\documentclass[12pt]{JHEP3}
\usepackage{amsmath, amsfonts,euscript,bbm}
\usepackage{epsfig, cite}
\usepackage{ifthen}
\usepackage{graphicx}
\usepackage{psfrag}

\usepackage{epstopdf}
\usepackage {amssymb}
\def\ba{\begin{eqnarray}}
\def\ea{\end{eqnarray}}
\newcommand\be{\begin{equation}}
\newcommand\ee{\end{equation}}
\newcommand\bse{\begin{subequations}}
\newcommand\ese{\end{subequations}}

\def\Tr{  \mbox{Tr}   }
\def\STr{  \mbox{STr}   }

\input{epsf.sty}

\title{{\Huge{M-flation}}:\\ {Inflation From Matrix Valued Scalar
Fields}}

\author{ Amjad Ashoorioon \footnote{amjad@umich.edu}\\
Michigan Center for Theoretical Physics, University of
Michigan, Ann Arbor, Michigan 48109-1040, USA}
\author{
Hassan Firouzjahi \footnote{firouz@physics.mcgill.ca}  \\
 School of Physics, Institute for Research in Fundamental Sciences (IPM),
\\ P .O. Box 19395-5531, Tehran, Iran}

\author{ M. M. Sheikh-Jabbari  \footnote{jabbari@theory.ipm.ac.ir}\\
 School of Physics, Institute for Research in Fundamental Sciences (IPM), \\ P .O. Box 19395-5531,Tehran, Iran}

\abstract{We propose an inflationary scenario, \emph{M-flation}, in
which inflation is driven by three $N\times N$ hermitian matrices
$\Phi_i,\ i=1,2,3$. The inflation potential of our model, which
is strongly motivated from string theory, is constructed from
$\Phi_{i}$ and their commutators. We show that one can consistently
restrict the classical dynamics to a sector in which the $\Phi_i$
are proportional to the $N\times N$ irreducible representations of
$SU(2)$. In this sector our model effectively behaves as an
N-flation  model with $3 N^2$ number of fields and the effective inflaton field has a super-Planckian field value. Furthermore, the  fine-tunings associated with unnaturally small couplings in the chaotic type inflationary scenarios are removed.
Due to the matrix nature of the inflaton fields
there are $3N^2-1$ extra scalar fields in the dynamics.
These have the observational effects such as production of iso-curvature
perturbations on cosmic microwave background.
Moreover, the existence of these extra scalars provides us with a
natural preheating mechanism and exit from inflation. As the
effective inflaton field can traverse super-Planckian distances in the
field space, the model is capable of producing a considerable amount
of gravity waves that can  be probed by future CMB polarization
experiments such as PLANCK, QUIET and CMBPOL.}


\preprint{IPM/P-2009/009\\ arXiv:0903.1481 [hep-th]}

\begin{document}

%%%%%%%%%%%%%%%%%%%%%%%%%%%%%%%%%%%%%%%%%%%%%%%%%%%%%%%%%%%%%%%%%%%%%%%%%%%%%%%%%%%%%%%%%%%%
\section{Introduction}

Recent observations, specially the five years Wilkinson Microwave
Anisotropy Probe (WMAP5) data \cite{Komatsu:2008hk} strongly support
inflation as the theory of early Universe and structure formation.
In their simplest forms, models of inflation are constructed from a
scalar field, the inflaton field, which is minimally coupled to
gravity. The potential is flat enough, so that a  period of
slow-roll inflation is achieved.  These simple models of inflation,
not only solve the problems associated with the standard big bang
cosmology such as the horizon problem, the flatness problem and the
monopole problem, but also generate quantum perturbations  which
seed the structures in the Universe. The slow-roll models of
inflation produce perturbations on cosmic microwave background (CMB)
which are almost scale-invariant, adiabatic and Gaussian  which are
in very good agreement with the current data. Theoretically, on the
other hand, inflation  still  remains a paradigm and one can
construct many inflationary models compatible with the current data.
There have been many attempts to embed inflation in string theory,
for a review see \cite{HenryTye:2006uv}.

Theories of multi-field inflation, in which one deals with more than
one scalar field, have also been studied \cite{Gordon:2000hv}, for a
review see \cite{Wands:2007bd}.  In the multiple field inflationary models
one can perform a rotation in the field space of scalar fields where
the inflaton field is evolving along a  trajectory while the remaining fields
are orthogonal to it.
These extra fields, like the inflaton
itself, have quantum fluctuations which once stretched to
super-Hubble scales can become classical and can therefore
contribute to the power spectrum of iso-curvature as well as
curvature perturbations, the details of which depends on the post
inflationary dynamics and the reheating scenario.  A specific
possibility in the multi-field inflation is the idea of assisted
inflation \cite{Liddle:1998jc}: while the potential is too steep for
an individual field to support inflation, the collective effect of a
large number of scalar fields leads to enough number of e-folds.
Similar idea was exploited in N-flation \cite{N-flation} and Cascade
inflation \cite{Becker:2005sg, Ashoorioon:2006wc} to obtain
inflation from several potentials that are individually too steep to
sustain inflation. In these models, although the ``effective''
inflaton field gets a super-Planckian field value for chaotic $m^{2
} \phi^{2}$ and $\lambda \phi^{4}$ inflationary scenarios, due to
the large number of fields, each physical field remains
sub-Plankian. Moreover, for the case of the $\lambda \phi^4$
inflationary theory this can be used to resolve problem of
unnaturally small coupling \cite{Kanti:1999vt,Kanti:1999ie}. Also
following the Lyth bound \cite{Lyth:1996im}, due to the large
excursions of the effective inflaton in comparison with $M_{P}$, in
these models one expects to obtain a considerable amount of gravity
waves.

In this work we take a different view and promote the inflaton
fields to general $N\times N$  hermitian  matrices and hence these
models will be called Matrix Inflation or \emph{M-flation} for
short. In this sense M-flation is a special case of multi-field
inflation. Working with matrices, besides the simple products of the
fields, we can also consider commutators of matrices. In our class
of M-flation models we consider three $N\times N$ matrices,
$\Phi_i,\ i=1,2,3$ and the potential is taken to be quadratic in the
$\Phi_i$ or in their commutator $[\Phi_i,\Phi_j]$. Therefore, in the
class of models we consider the potential term for $\Phi_i$ can have
three types of terms: $\Tr\left [\Phi_i,\Phi_j \right ]^2$, $\Tr \,
\epsilon_{ijk}\Phi_i \left[ \Phi_j,\Phi_k \right]$ and $\Tr \, \Phi_i^2$. As we will
argue, this class of potentials is well-motivated from string theory
and  brane dynamics.

As we will see despite the simple form of the potential constructed
from these matrices and their commutators our model has a rich
dynamics. We argue that M-flation can solve the fine-tunings
associated with standard chaotic inflationary scenarios.
Furthermore, like any multi-field inflation model, there would be
iso-curvature perturbations as well as the usual adiabatic
perturbations. This can have significant observational consequences
for the CMB observations \cite{Bean:2006qz,Komatsu:2008hk}.
Moreover, we argue that our model has an embedded efficient
preheating mechanism.
% and a natural mechanism of exit from inflation.

The outline of the paper is as follows. In  section
\ref{setup-section}, we provide the setup through introducing the
action and show that, with the appropriate initial conditions, for
the sector in which the $\Phi_i$ fields fall into irreducible
$N\times N$ representations of $SU(2)$, the theory effectively and
at the classical level, behaves like a single field inflation. In
section \ref{Inflation-in-M-flation}, we discuss simple models of
inflation such as chaotic, symmetry breaking and inflection point
inflation which are constructed from our M-flation. In section
\ref{mass-spectrum-section}, we work out the mass spectrum of  the
remaining $3N^2-1$ scalar fields, the iso-curvature modes, which are
not classically turned on during inflation (these fields will be
generically called $\Psi$-modes). In sections \ref{power-spectrum}
and \ref{preheating-section} we consider quantum mechanical
excitations of $\Psi$-modes and their effects. In section
\ref{power-spectrum}, we compute the power spectrum of fluctuations
of the inflaton and the $3N^2-1$ iso-curvature modes. In section
\ref{preheating-section}, we focus on quantum mechanical creation of
the sub-Hubble  modes. This particle creation takes  energy away
from the inflaton field.  We show that our model naturally contains
this mechanism \cite{Traschen:1990sw} which is the basis of the
preheating scenarios \cite{preheating-I,preheating-II}. In section
\ref{stringtheory}, we give string theory motivations behind the
M-flation models, arising from dynamics of multiple D3-branes in
specific flux compactifications. The last section is devoted to
discussion and outlook. In the Appendix we have gathered some
technical details of slow-roll inflation.

%%%%%%%%%%%%%%%%%%%%%%%%%%%%%%%%%%%%%%%%%%%%%%%%%%%%%%%%%%%%%%%%%%%%%%%%%%%%%%%%%%%%%%%%%%%%
\section{M-flation scenario, the setup}\label{setup-section}

As explained before, we start with an inflationary setup in which
the inflaton fields are taken to be $N\times N$  non-commutative
hermitian matrices. We start from the following action %
\ba \label{action}%
 S=\int d^{4} x \sqrt {-g} \left(\frac{ M_{P}^{2} }{2} R - \frac{1}{2}
\sum_{i} \Tr  \left( \partial_{\mu} \Phi_{i} \partial^{\mu} \Phi_{i}
\right) - V(\Phi_{i}, [ \Phi_{i}, \Phi_{j}] ) \right) \, , %
\ea %
where the reduced Planck mass is $M_{P}^{-2}= 8 \pi G$ with $G$
being the Newton constant and the  signature of the metric is
$(-,+,+,+)$. Note that the $i$-index counts the number of matrices
and is not denoting the matrix elements; the matrix element indices
are suppressed here. Also, $V$ represents our potential constructed
from the $N \times N$ matrices $\Phi_{i}$ and their commutators $[
\Phi_{i}, \Phi_{j}] $. The kinetic energy for $\Phi_{i}$ has the
standard form and it is assumed that the $\Phi_{i}$ matrices are
minimally coupled to gravity.

As we will discuss in section \ref{stringtheory}, the potential
$V(\Phi_{i}, [ \Phi_{i}, \Phi_{j}])$ can be motivated from dynamics
of branes in string theory where up to leading terms in
$\Phi_{i}$ and $[ \Phi_{i}, \Phi_{j}] $, in specific string theory backgrounds, the potential takes the form%
\ba\label{The-Potential}%
V= \Tr  \left( - \frac{\lambda}{4}  [ \Phi_{i},
\Phi_{j}] [ \Phi_{i}, \Phi_{j}] +\frac{i \kappa}{3} \epsilon_{jkl}
[\Phi_{k}, \Phi_{l} ] \Phi_{j} +  \frac{m^{2}}{2}  \Phi_{i}^{2}
\right),%
\ea%
where $i=1,2,3$ and hence we are dealing with $3N^2$ real scalar
fields. The $\Tr$ is over $N\times N$ matrices, and here and below
the summation over repeated $i,j$ indices is assumed. $ \lambda$ is
a dimensionless number while $\kappa$ and $m$ are constants with
dimensions of mass. We take $\lambda$, $\kappa$ and $m^2$ to be
positive. Note that the potential \eqref{The-Potential} is quadratic
in powers of $[\Phi_i,\Phi_j]$ and $\Phi_i$. The action
\eqref{action} together with the potential \eqref{The-Potential} are
invariant under $U(N)$ (acting on the matrices) and $SU(2)$ (acting
on $i,j$ indices) which are both global symmetries.

Starting with an FRW background
\be%
ds^{2} = -dt^{2} + a(t)^{2} d   {\bf x}^{\, 2} \, ,
\ee%
the equation of motions are%
 \bse\label{eom}\begin{align}%
&H^{2}= \frac{1}{3 M^2_{P}} \left( - \frac{1}{2} \Tr
\left(
\partial_{\mu} \Phi_{i} \partial^{\mu} \Phi_{i}  \right) +
V(\Phi_{i}, [ \Phi_{i}, \Phi_{j}] ) \right) \\ &\ddot \Phi_{l} + 3 H
\dot \Phi_{l} + \lambda   \left[ \Phi_{j}, \, [\Phi_{l}, \Phi_{j}]
\,  \right] + i \, \kappa \, \epsilon_{l j k } [ \Phi_{j}, \Phi_{k}]
+ m^{2} \Phi_{l} =0 \, ,\\
&\dot H=-\frac{1}{2M^2_{P}}\ \sum_{i} \Tr
\partial_{\mu} \Phi_{i} \partial^{\mu} \Phi_{i}\ ,%
\end{align}\ese%
 where $H= \dot a/a$ is the Hubble expansion rate.

%%%%%%%%%%%%%%%%%%%%%%%%%%%%%%%%%%%%%%%%%%%%%%
\subsection{Truncation to the $SU(2)$ sector}

As discussed, $\Phi_{i}$ are $N \times N$ matrices and we are hence
dealing with $3N^2$ real scalar fields which are generically coupled to
each other. This makes the analysis of the model in the most general
form very difficult, if not impossible. Noting the specific form of the
the potential \eqref{The-Potential} and that $i,j$ indices are
running from $1$ to $3$, there is the  possibility of consistently
restricting the classical dynamics to a sector in which we are
effectively dealing with a single scalar field. This sector, which will
be called the $SU(2)$ sector,
is obtained for matrix configurations of the form%
\ba \label{phiJ}%
\Phi_{i} =
\hat \phi(t) J_{i}\ , \quad \quad i=1,2,3,%
\ea%
where $J_{i}$ are the basis for the $N$ dimensional irreducible
representation of the $SU(2)$ algebra%
\ba\label{J}%
 [ J_{i}, J_{j} ]=  i \, \epsilon_{ijk} J_{k} \ , \qquad \Tr (J_{i} \,
J_{j})= \frac{N}{12} ( N^{2}-1 ) \, \delta_{ij} \, .%
\ea%
Since both $\Phi_{i}$ and $J_{i}$ are hermitian, we conclude that
$\hat \phi$ is a real scalar field.

Plugging these into the action (\ref{action}), we obtain \ba S= \int
d^{4} x \sqrt {-g} \left[ \frac{M_{P}^{2}}{2} R+ \Tr J^{2}  \left( -
\frac{1}{2}  \partial_{\mu} \hat \phi  \partial^{\mu} \hat \phi
-\frac{\lambda}{2}  \hat \phi^{4}  + \frac{2 \kappa}{3} \hat
\phi^{3} - \frac{m^{2}}{2} \hat \phi^{2} \right) \right] \, , \ea
where $\Tr J^{2} = \sum_{i=1}^{3} \Tr (J_{i}^{2})   = N(N^{2}
-1)/4$.

Interestingly enough, this represents the action of  chaotic
inflationary models with a non-standard kinetic energy. Upon the
field redefinition%
\ba \label{phi-scaling}%
 \hat \phi = \left(  \Tr
J^{2}   \right)^{-1/2} \phi = \left[ \frac{N}{4}(N^{2}-1)
\right]^{-1/2} \, \phi \, , %
\ea%
the kinetic energy for the new field
$\phi$ becomes standard, while the potential for it becomes %
\ba\label{Vphi}%
V_0(\phi)= \frac{\lambda_{eff}}{4} \phi^{4} -
\frac{2\kappa_{eff}}{3} \phi^{3} + \frac{m^{2}}{2} \phi^{2} \, , %
\ea%
where%
\ba \label{lameff}%
\lambda_{eff} = \frac{2 \lambda}{\Tr J^{2}} = \frac{8 \lambda}{ N
(N^{2}-1)}  \ , \quad \kappa_{eff} = \frac{ \kappa}{\sqrt{\Tr J^{2}}
} = \frac{2 \,
\kappa}{\sqrt{N(N^{2}-1)}} %
\ea%

%%%%%%%%%%%%%%%%%%%%%%%%%%%%%%%%%%%%%%%%%%%%%%
\subsection{Consistency of the truncation to the $SU(2)$
sector}\label{consistent-truncation-section}

The $SU(2)$ sector seems to be a special sector of the M-flation
action in which the theory becomes tractable and very simple.
However, we need to make sure that this truncation to the $SU(2)$
sector is indeed consistent with the classical dynamics of the model
and that we can consistently turn off the other $3N^2-1$ fields. In
order to do this, we define%
\be\label{Psi-i}%
\Psi_i=\Phi_i-\hat \phi J_i%
\ee%
where, as before, \[ \hat \phi= \frac{4}{N(N^2-1)}\Tr(\Phi_i J_i)\ ,
\]
and hence $\Tr(\Psi_i J_i)=0$. In other words, $\Psi_i$ is defined
such that it has no components along $J_i$. Using $[J_i,
J_j]=i\epsilon_{ijk}J_k$, we may rewrite the potential
\eqref{The-Potential} in terms of $\hat \phi$ and  $\Psi_i$ as
\be\label{pot-psi-phi}%
V=V_0(\hat \phi)+ V_{(2)}(\hat \phi, \Psi_i)
\ee%
where $V_0$, after the field re-definition \eqref{phi-scaling}, is
given by \eqref{Vphi}.
% \%
%\be\begin{split}%
%V_0(\hat \phi) &=\Tr J^{2}  \left( \frac{\lambda}{2}  \hat \phi^{4}
%- \frac{2 \kappa}{3} \hat \phi^{3}+ \frac{m^{2}}{2} \hat \phi^{2}
%\right)\cr%
%&= \frac{\lambda_{eff}}{4} \phi^{4} -
%\frac{2\kappa_{eff}}{3} \phi^{3} + \frac{m^{2}}{2} \phi^{2} \,%
%\end{split}
%\ee%

Using \eqref{J}, \eqref{Psi-i} and the fact that $\Tr(\Psi_i
J_i)=0$, one can show that $V_{(2)}$ does not have any linear terms
in $\Psi_{i}$, i.e.
%
%$V_{(2)}(\hat \phi,\Psi_i)$  does not have any linear
%terms in $\Psi_i$, i.e. %
\ba\label{V(2)-Psi2}%
V_{(2)}(\hat\phi, {\Psi_i=0})=0\ ,\qquad \left(\frac{\delta
V_{(2)}}{\delta\Psi_i}\right)_{\Psi_i=0}=0. %
\ea%
This leads to  the important result that the $\phi$ field does not
source the $\Psi_i$ fields. Explicitly, in the equations of motion
(\ref{eom}b) if we start with the initial conditions $\Psi_i=0,\
\dot\Psi_i=0$ and $\hat \phi\neq 0$, $\Psi_i$ will always remain
zero and will hence not contribute to the classical background
inflationary dynamics at all. This means that we are consistent in
considering  $\phi$ as the sole field driving the inflation. The
remaining $\Psi_i$ modes, however, as we shall see in next sections,
will be excited at the level of perturbations. So, we have
$3N^{2}-1$ $\Psi$ modes besides the background $\phi$ inflationary
field. In the language of \cite{Gordon:2000hv} this corresponds to a
straight inflationary trajectory in the field space of $3N^{2}$
scalar fields, where the inflationary trajectory is along the $\phi$
direction and there would be $3N^{2}-1$ iso-curvature perturbations
perpendicular to the adiabatic trajectory. We should stress that
this result will not remain valid if initially $\Psi_i$ fields are
also turned on.

One may wonder about the special role of the $SU(2)$ generators
$J_i$ among other $N\times N$ matrices in our analysis and whether a
similar reduction to  a sector other than the $SU(2)$ sector is also
possible. Traces of the fact that $SU(2)$ sector is special is
already built in the initial construction of the potential
\eqref{The-Potential}  with three $\Phi_i$,  $i=1,2,3$ (3 is the
dimension of the $SU(2)$ algebra) and that in the cubic term in the
action the structure constant of $SU(2)$ $\epsilon_{ijk}$ appears.
To see this more explicitly, let us consider the more generic
decomposition for $\Phi_i$
matrices%
\be\label{Phi-Upsilon-decomp}%
\Phi_i=\Upsilon_i+\Xi_i\ ,%
\ee%
such that $\Tr(\Upsilon_i\Xi_i)=0$, we take both $\Upsilon_i$ and
$\Xi_i$ to be hermitian. The potential will again have two parts,
$V=V_0(\Upsilon_i)+V_{(1)}(\Upsilon, \Xi)$, where
$V_0(\Upsilon_i)=V(\Xi_i=0)$ and%
\[
V_{(1)}(\Upsilon_i,\Xi_i)=\Tr
\left[\biggl(-\lambda[\Upsilon_i,[\Upsilon_i,\Upsilon_k]]+i\kappa\epsilon_{ijk}[\Upsilon_i,\Upsilon_j]\right)\Xi_k\biggr]+{\cal
O}(\Xi^2)\ .%
\]%
In order the $\Upsilon$-sector to decouple, the above expression in
the bracket should vanish for any $\Xi_i$. As explained above, if
$\Upsilon_i$ is proportional to $J_i$ this condition obviously
holds. In general, however, this can happen if the
$\Upsilon$-dependent term in parenthesis is proportional to
$\Upsilon_k$.  This condition can be satisfied if
$[\Upsilon_i,\Upsilon_j]=f_{ijk}\Upsilon_k$ for some functions
$f_{ijk}$ which means  that the three $\Upsilon_i$ matrices
 should form a Lie-algebra of dimension three. The only
non-trivial solution is then $\Upsilon_i$ forming an $SU(2)$
algebra.\footnote{There is also the trivial choice of $f_{ijk}=0$,
corresponding to choosing three Abelian subgroups of $U(N)$. Working
with commuting matrices, however, kills all the interesting
inflationary dynamics.} Note, however, that the representation for
the $SU(2)$ algebra is not fixed by this requirement and
$\Upsilon_i$ could also form reducible $N\times N$ representation of
$SU(2)$.

Motivated by this unique property of the $SU(2)$ sector, from now on
we assume that the background inflationary trajectory is along the $
\phi$ direction, while the other $3N^{2}-1$ fields are only excited
at the level of (quantum) perturbations.

%%%%%%%%%%%%%%%%%%%%%%%%%%%%%%%%%%%%%%%%%%%%%%%%%%%%%%%%%%%%%%%%%%%%%%%%%%%%%%%%%%%%%%%%%%%%
\section{Various inflation models resulting from M-flation
scenario}\label{Inflation-in-M-flation}%

The $SU(2)$ sector, which is governed by the potential \eqref{Vphi},
depending on the values of the parameters $\lambda_{eff}$,
$\kappa_{eff}$ and $m^2$, leads to different inflationary models.
These models have been studied in the literature, which we will
review below. The important advantage of our M-flation scenario, as
we will show, is the scaling properties between the original
parameters of the model appearing in the action and the effective
dressed parameters appearing in \eqref{Vphi}, which allow us to
remove the unnaturalness and fine-tuning of these parameters.

%%%%%%%%%%%%%%%%%%%%%%%%%%%%%%%%%%%%%%%%%%%%%%
\subsection{Chaotic inflation}
\label{chaotic}

If we set $\lambda=\kappa=0$, we obtain the simple quadratic chaotic
potential and in this case our model is nothing but an N-flation
model \cite{N-flation} with $3N^2$ fields. To fit the CMB
observations and  obtain right number of e-foldings, one needs that
$m \sim 10^{12}$ GeV and a super-Planckian  field variation for
effective inflaton field $\phi$ during inflation, $\Delta \phi \sim
10 M_{P}$. In the context of effective field theory this may sound
problematic. However, as in the N-flation model \cite{N-flation},
note that $\hat \phi$ is our physical field and $\Delta \hat \phi
\sim \Delta \phi /N  $. For a sufficiently large value of
$N$ one can arrange that $\Delta \hat \phi  \ll M_{P}$ and one can
avoid the problem with super-Planckian field values.

 On the other hand, if $m=\kappa=0$,  we obtain the
quartic $\lambda_{eff}\phi^4/4$ chaotic inflation. To fit the COBE
normalization and obtain right number of e-folds, one requires
$\lambda_{eff} \sim 10^{-14}$ and $\Delta \phi \gtrsim 10 M_{P}$. In
the context of a single scalar field, these are viewed as severe
fine-tunings in the model. In our case, however, we see that both of
these can be relaxed. Assuming that $\lambda \sim 1$ dictated by the
naturalness of the theory, to obtain the above value for
$\lambda_{eff} $  one needs to have $N \sim 10^{5}$. In the context
of string theory studied in section \ref{stringtheory}, where $N$ is
viewed as the number of coincident branes, this is easily achieved
in light of recent developments in the flux compactification. With
this value of $N$ for the physical field $\hat \phi$ one obtains
$\Delta \hat \phi \sim 10^{-7} M_{P}$ and  can hence safely bypass
the problem of super-Planckian excursion of the field. In both of
these examples, inflation ends when $\hat \phi(t) \rightarrow 0$
which means that the matrices $\Phi_{i}$ commute with each other.

In both examples, and for next two examples below, due to
super-Planckian value of $\Delta \phi$ during inflation, a
considerable amount of gravity waves can be produced which can be
detected in future gravity wave observations such as PLANCK
\cite{PLANCK, Efstathiou:2009xv}, CMBPOL \cite{Baumann:2008aq} and QUIET \cite{quiet}.

%%%%%%%%%%%%%%%%%%%%%%%%%%%%%%%%%%%%%%%%%%%%%%%%%%%%%%%%%%%%%%%%%%%%%%%%%%%%%%%%%%%%%%%%%%%%

\subsection{Symmetry breaking inflation}\label{SBI} %

Now consider the general case where none of the coefficients in
$V_0(\phi)$ is zero. We study two interesting example here. The
first example is when $V_0(\phi)$ is positive definite and has two
degenerate minima. This corresponds to $\kappa
= 3\,  m \sqrt {\lambda}/2$ and the potential has the form %
\ba\label{Vsusy}%
V_0=  \frac{\lambda_{eff}}{4} \phi^{2}  \, (\phi- \mu)^{2}%
\ea%
 where $\mu \equiv \sqrt 2 m/\sqrt{\lambda_{eff}}$. As mentioned, the
potential has global minima at $\phi=0$ and $\phi=\mu$. In the
language of the brane construction (\emph{cf.} section
\ref{stringtheory}), the minimum at $\phi=\mu$ corresponds to
super-symmetric vacua when $N$ D3-branes blow up into a giant
D5-brane in the presence of background RR field $C^{(6)}$ which in
our notation is represented by $\kappa$. The minimum at $\phi=0$, on
the other hand, corresponds to the trivial solution when matrices
become commutative. If we allow the field $\phi$ to take negative
values, then the potential is symmetric under $\phi \rightarrow
-\phi + \mu$ and it represents the standard double well potential,
justifying the name ``symmetry breaking'' inflation.

Potential (\ref{Vsusy}) is well studied in the literature and is in
the form  of  symmetry breaking potential and ``hilltop inflation''
\cite{Boubekeur:2005zm, Freese:1990rb}. In Appendix  \ref{Appen-A}
we briefly look at the predictions from this potential compared to
WMAP5 data. It is assumed that inflation lasts for 60 number of
e-folds, $N_{e}=60$, and the scalar spectral index, $n_{s}=0.96$,
from WMAP5 central value. Furthermore, the COBE normalization is set
to $\delta_{H} \simeq 2.41 \times 10^{-5}$.

Depending on the initial value of the inflaton field, $\phi_{i}$,
the inflationary period is divided in three categories.

%%%%%%%%%%%%%%%%%%%%%%%%%%%%%%%%%%%%%%%%%%%%%%

\subsubsection*{{\bf (a)}\ {$\ \phi_{i}> \mu$}}\label{phi-grtr-mu}
Suppose inflation starts when $ \phi_{i}> \mu$. With $N_{e}=60, \delta_{H} \simeq 2.41 \times 10^{-5}$ and $n_{s}=0.96$, one obtains%
 \ba \label{case1}%
 \phi_{i} \simeq 43.57
M_{P} \quad , \quad \phi_{f} \simeq 27.07 M_{P} \quad , \quad \mu
\simeq 26 M_{P}.%
 \ea%
  and
\begin{equation}\label{phi>mu}
% \nonumber to remove numbering (before each equation)
   \lambda_{eff}\simeq 4.91 \times 10^{-14}, \quad m\simeq 4.07\times 10^{-6} M_{P}, \quad \kappa_{eff}\simeq 9.57 \times 10^{-13} M_{P}.
\end{equation}

%%%%%%%%%%%%%%%%%%%%%%%%%%%%%%%%%%%%%%%%%%%%%%
\subsubsection*{{\bf (b)}$\ \ \mu/2 <\phi_{i}< \mu$}
This is an example of hill-top inflation  when the inflaton field is
between the local maximum at $\phi=\mu/2$ and the ``supersymmetric
minimum'' at $\phi=\mu$.  To fit the above observational constraints
one obtains \ba \label{case2} \phi_{i} \simeq 23.5 M_{P} \quad ,
\quad \phi_{f} \simeq 35.03 M_{P} \quad , \quad \mu \simeq 36 M_{P}.
\ea and
\begin{equation}\label{phi<mu}
% \nonumber to remove numbering (before each equation)
   \lambda_{eff}\simeq 7.18\times 10^{-14}, \quad m\simeq 6.82\times 10^{-6} M_{P}, \quad \kappa_{eff} \simeq 1.94\times 10^{-12} M_{P}.
\end{equation}

%%%%%%%%%%%%%%%%%%%%%%%%%%%%%%%%%%%%%%%%%%%%%%
\subsubsection*{{\bf (c)}$\ \ 0 <\phi_{i}< \mu/2$}
Due to symmetry $\phi \rightarrow -\phi + \mu$ this inflationary
region has the same properties as $ \mu/2 <\phi_{i}< \mu$ above.  If
we allow for negative values of $\phi$, then the inflationary
prediction for $\phi<0$ region is the same as in region
\ref{phi-grtr-mu} above.

In all these examples, to fit the COBE normalization, one obtains $N\sim 10^{5}$ and $\Delta \hat \phi \sim 10^{-7} M_{P}$ during inflation  so the issues with super-Planckian field range is resolved.

%%%%%%%%%%%%%%%%%%%%%%%%%%%%%%%%%%%%%%%%%%%%%%%%%%%%%%%%%%%%%%%%%%%%%%%%%%%%%%%%%%%%%%%%%%%%
\subsection{Inflection point inflation}
\label{saddle}

Another class of models widely studied in the literature coming from
the potential (\ref{Vphi}) is when the potential has an inflection
point \cite{Allahverdi:2006iq}. This happens when $\kappa = \sqrt{2
\lambda} \, m$ or equivalently, $\kappa_{eff} = m
\sqrt{\lambda_{eff}}$. Denoting the inflection point by $\phi_{0}$,
the potential near $\phi_{0}$ is approximately given by%
\ba%
V(\phi)\simeq V(\phi_{0}) + \frac{1}{3!} V'''(\phi_{0})
(\phi-\phi_{0})^{3} \ea where \ba V(\phi_{0}) = \frac{m^{2}}{12}
\phi_{0}^{2} \quad , \quad V'''(\phi_{0}) = \frac{2 m^{2}}{
\phi_{0}} \, .%
\ea%

The CMB observables  are given by \cite{Allahverdi:2006iq}%
\ba%
n_{s} \simeq 1- \frac{4}{N_{e}}  \quad , \quad
\delta_{H} \simeq \frac{2}{5 \pi} \frac{\lambda_{eff} M_{P}}{m} N_{e}^{2} \, .%
\ea%

Choosing $N_{e}=60$, one obtains $n_{s}\simeq 0.93$ which is within
$2\sigma$ error bar of WMAP5 but is somewhat to its lower end. One
may add small modification to the coefficient of potential
\cite{Allahverdi:2006iq} such that the potential around the
inflection point is slightly modified. This in turn can result in a
higher value of $n_{s}$. On the other hand, from COBE normalization,
one obtains \ba\label{lambda-m} \lambda_{eff} \sim 10^{-8}
\frac{m}{M_{P}} \, . \ea In a conservative limit that $m\lesssim
M_{P}$, this yields $\lambda_{eff} \lesssim 10^{-8}$. Starting with
$\lambda\sim 1$, this corresponds to $N\gtrsim 10^{3}$.

If we look into the amplitude of the gravity wave, determined by
quantity $r$  which is the ratio of gravitational perturbation
amplitude to scalar perturbation amplitude, one obtains \ba r= 8
M_{P}^{2} \left(\frac{V'}{V}\right)^{2}= \frac{2}{9}
\left(\frac{m}{\sqrt{\lambda_{eff}}} \right)^{6} N_{e}^{-4} \, . \ea
Combining this with $n_{s}$ and $\delta_{H}$, we obtain \ba
\lambda_{eff}= (\frac{9\, r}{32})^{1/3} \left(\frac{5 \pi}{8}
\delta_{H} \right)^{2} (1- n_{s})^{8/3} \, . \ea The upper bound,
$r<0.22$, from WMAP5 implies that $\lambda_{eff} \lesssim 10^{-13}$
and $N \gtrsim 10^{5}$ which is stronger than the bound above.

%%%%%%%%%%%%%%%%%%%%%%%%%%%%%%%%%%%%%%%%%%%%%%%%%%%%%%%%%%%%%%%%%%%%%%%%%%%%%%%%%%%%%%%%%%%%
\section{Mass spectrum of $\Psi_i$ modes in
M-flation}\label{mass-spectrum-section}

Having studied the $SU(2)$ sector and the resulting inflationary
models, we review our model, noting that the other $3N^2-1$ fields
encoded in $\Psi_i$, although not contributing to the classical
inflationary dynamics, do have quantum fluctuations and  will hence
affect the cosmological perturbation analysis. To compute these effects we need to have the
mass spectrum of the $3N^2-1$ modes coming from the $\Psi_i$.

To this end, starting from \eqref{Psi-i} we expand the action up to
the second order in $\Psi_i$. Given the orthogonality condition,
$\Tr(\Psi_i J_i)=0$, the kinetic term readily takes the standard
form $\frac{1}{2}  \Tr(\partial_\mu \Psi_i\partial^\mu\Psi_i)$.
After a slightly lengthy
but straightforward  computation the potential to second order in $\Psi_i$
is obtained as%
\be\label{V-Psi-2}%
V_{(2)}=  \Tr \left[
\frac{\lambda}{2} \hat  \phi^2\
\Omega_i  \Omega_i+\frac{m^2}{2}\
 \Psi_i\Psi_i+\left(-\frac{\lambda}{2}\hat \phi^2+\kappa \hat  \phi\right)
\Psi_i\Omega_i\ \right] \,
\ee%
where%
\be\label{Omega-Psi}%
\Omega_k\equiv i\epsilon_{ijk}[J_i, \Psi_j]\ .%
\ee

>From the above form we see that if we have the eigenvectors
(eigen-matrices) of the $\Omega_i$ we can compute the spectrum of
$\Psi_i$ in terms of $\hat \phi$-field (to be viewed as the
inflaton). Finding the eigenvectors of $\Omega_i$ is mathematically
the same problem as finding the vector spherical harmonics. (For a
detailed discussion see e.g. \cite{Dasgupta:2002hx}, section 5.2.)
If we
denote the $\Omega$ eigenvalues by $\omega$, i.e.%
\be\label{Omega-eigenvalue}%
\Omega_i=\omega \Psi_i\ ,%
\ee%
we obtain
\be\label{V-Psi-2-mu}%
V_{(2)}= \left(\frac{\lambda_{eff}}{4} \phi^2 (\omega^2-\omega)+ \kappa_{eff} \, \omega \,
\phi+\frac{m^2}{2} \right)\
\Tr \, \Psi_i\Psi_i\ .%
\ee%
If we have the possible values of $\omega$ we can read off the
effective ($\phi$-dependent) mass of the $\Psi_i$ modes%
\be%
\label{massM}
\begin{split}%
M^2 &=\frac{\lambda_{eff}}{2} \phi^2 (\omega^2-\omega)+
2\kappa_{eff} \, \omega
 \phi+m^2 \cr%
&= V''_0 (\omega+1)^2-\frac{V'_0}{\phi}
(4\omega+3)(\omega+2)+\frac{6V_0}{\phi^2} (\omega+1)(\omega+2)\ ,
\end{split}
\ee%
where $V_0'$ and $V_0''$ denote the first and second derivatives of
the potential $V_0$ with respect to the inflaton $\phi$. For
later convenience it is also useful to write the expression for the
mass during inflation in terms of the Hubble expansion rate and the
slow-roll parameters $\epsilon$ and $\eta$%
\be\label{M-slow-roll}%
\frac{M^2}{3H^2}=\left[\eta (\omega+1)^2- sgn(V_{0}')  \sqrt{2\epsilon}\
\frac{M_{P}}{\phi}(4\omega+3)(\omega+2)+6\frac{M_{P}^2}{\phi^2}(\omega+1)(\omega+2)\right]  \, ,%
\ee%
where $sgn(V_{0}')$ represents the sign of $V_{0}'$ and as usual
the slow-roll parameters are defined by%
\ba\label{epsilon-eta}%
 \epsilon = \frac{M_{P}^{2}}{2} \left( \frac{V_{0}'}{V_{0}}  \right)^{2}
 \ , \qquad
 \eta = M_{P}^{2} \frac{V_{0}''}{V_{0}}.
 \ea

Following the analysis of \cite{Dasgupta:2002hx}, we find that $\omega$
can take three values:
\begin{itemize}%
\item \emph{`` The zero modes'' $\omega=-1$}. This happens for modes of
the form%
\[ \Psi_i=[J_i, \Lambda],%
\]%
with $\Lambda$ being an arbitrary traceless matrix. (Note that
$\Lambda\propto {\bf 1}_N$ matrix is not a zero mode.) For these
modes the
expression for the mass simplifies to%
\be\label{Mass-zero-mode}%
%\begin{split}%
{M^2}%%{3H^2}\  sgn(V_{0}') \sqrt{2\epsilon} \, \frac{M_{P}}{\phi}
=\frac{V_{0}'}{\phi}\ .%
\ee%
Therefore, at the minimum values for $ \phi$ where $V_0'$ vanishes
these modes become massless. (This justifies the name ``massless
modes''.)

Noting that $\Lambda$ is an arbitrary matrix there are $N^2-1$ of
such modes, all with the same mass.

\item \emph{``The $\alpha$ modes''}: $\omega=-(l+2)$, $l\in
\mathbb{Z},\ 0\leq l\leq N-2$, with the mass%
\be\label{Mass-alpha-mode}%
%\frac{M^2_l}{3 H^{2}}= %\eta \,  l^{2} - sgn(V_{0}') \sqrt{2\epsilon}
%\, \frac{M_P}{\phi}  (4 l +1 )(l-1) + 6\frac{M_P^2}{\phi^2} \,  l (l-1)  \ .%
M_{l}^{2}= \frac{\lambda_{eff}}{2} (l+2) (l+3) \phi^{2} - 2
\kappa_{eff} (l+2) \phi + m^{2} \, .
\ee%
Each mode for a given $l$ has a multiplicity of $2l+1$ and therefore, there are $(N-1)^2$ $\alpha$-modes.
Of course, $l=0$ $\alpha$-mode is nothing more that the adiabatic mode which is the fluctuations of the inflaton field along the SU(2) sector itself.
Therefore, there are $(N-1)^2-1$ ``isocurvature'' $\alpha$-modes. 

\item \emph{``The $\beta$ modes''}: $\omega=l-1$, $l\in
\mathbb{Z},\ 1\leq l\leq N$, with the mass%
\be\label{Mass-beta-mode}%
%\frac{M^2_l}{3 H^{2}}= \eta (l+1)^{2} - sgn(V_{0}') \sqrt{2\epsilon}
%\, \frac{M_P}{\phi}  (4 l +3 )(l+2) + 6\frac{M_P^2}{\phi^2} (l+1)(l+2) \ .%
M_{l}^{2}= \frac{\lambda_{eff}}{2} (l-2)(l-1) \phi^{2}  + 2
\kappa_{eff} (l-1)  \phi + m^{2} \, . \ee Each mode for a given $l$
has a multiplicity of $2l+1$. Therefore, there are $(N+1)^2-1$ of
$\beta$-modes.
\end{itemize}

As expected, there are altogether $3N^2-1$ zero, $\alpha$, and $\beta$
isocurvature modes.  For the zero modes where $\omega=-1$, one observes that
$M^{2}/3H^{2} \sim \sqrt \epsilon M_{P}/\phi \sim 0.01$. So there
are $N^{2}-1$ of such light zero modes. For $\alpha$ and $\beta$
modes, only those with $l \lesssim  \epsilon^{-1/2} , \eta^{-1/2}
\sim 10$ are light, while the  modes with higher values of $l$ are
heavy.

%%%%%%%%%%%%%%%%%%%%%%%%%%%%%%%%%%%%%%%%%%%%%%%%%%%%%%%%%%%%%%%%%%%%%%%%%%%%%%%%%%%%%%%%%%%%

\section{Power spectra in the presence of $\Psi_i$ modes}\label{power-spectrum}%

With the mass spectrum for $\Psi_i$ modes computed in the previous
section we can compute the power spectra of the adiabatic and the
iso-curvature perturbations. Here  we are dealing with a $3N^{2}$
real scalar field inflationary system. Our inflationary background
is along the $\phi$ direction. The remaining $3 N^{2}-1$ scalars are
frozen classically during inflation and are excited only  quantum
mechanically. Correspondingly, the modes are classified as the
adiabatic perturbation, the one which is tangential to the classical
inflationary background, and the iso-curvature modes, the  $3
N^{2}-1$ perturbations which are perpendicular to the background
inflationary trajectory.  In this respect our model is similar to
the model studied in \cite{Byrnes:2005th} where the potential has an
$O(N)$ symmetry such that the inflaton field is the radial direction
while the remaining $N-1$ angular directions are iso-curvature
perturbations.

The formalism to calculate the adiabatic
and iso-curvature (entropy) power spectra  was systematically developed  in \cite{Gordon:2000hv}. Here we shall repeat those analysis for our system of $3N^{2}$
scalar fields.

%%%%%%%%%%%%%%%%%%%%%%%%%%%%%%%%%%%%%%%%%%%%%%
\subsection{Linear perturbations}

The perturbed metric in the longitudinal gauge is
\ba
\label{longitudinal}
ds^{2} = - (1+ 2 \Phi) dt^{2} + a(t)^{2} (1- 2 \Phi) d {\bf x}^{2}
\ea
where $\Phi(t, {\bf x})$ is the gravitational potential and should not be mistaken with
$\Phi_{i}$ which represents our matrix fields. Similarly, the linear perturbed scalar
fields are $\delta \phi(t, {\bf x})$ and $\Psi_{i}(t, {\bf x})$.

We find it is much easier to work with the $\Psi_{i}$ modes with
definite mass spectrum. As discussed in the previous section, these
fall into three classes denoted by $\Psi_{r, lm}$ where
$r=0,\alpha,\beta$ stand, respectively, for zero-mode, $\alpha$-mode and $\beta$-mode
 and $m$ is running from one to $D_{r, l}$, the degeneracy factor of each $\Psi_{r, lm}$ mode, with $D_0=N^2$  and $ \ D_{\alpha, l}=
D_{\beta, l}=2l+1$. In this notation, the Lagrangian  for the scalar
fields with potential $V= V_{0}(\phi) + V_{(2)} (\phi, \Psi_{i})$
from
(\ref{V-Psi-2-mu}) is%
\ba \label{S2}%
{\cal L}= -\frac{1}{2} \partial_{\mu} \phi  \partial^{\mu} \phi -
\frac{1}{2} \partial_{\mu} \Psi^\star_{r,lm}  \partial^{\mu}
\Psi_{r,lm} - V_{0}(\phi) - \frac{1}{2} M^{2}_{r,l}(\phi)
\Psi^\star_{r,lm}  \Psi_{r,lm}  \, , %
\ea%
where summation over repeated $r$ and $l,m$ indices is assumed.

 Define the gauge invariant Mukhanov-Sasaki variable $Q_{\phi}$
\ba Q_{\phi} \equiv \delta \phi + \frac{\dot \phi}{H} \Phi \, . \ea
The perturbed Klein-Gordon equations for $\phi$ and $\Psi_{r,lm}$
are \ba \label{KG} \ddot Q_{\phi} + 3 H \dot Q_{\phi} +
\frac{k^{2}}{a^{2}} Q_{\phi} + \left[ V_{0 \, , \phi \phi} -
\frac{1}{a^{3} M_{P}^{2}}
{\left(   \frac{a^{3}}{H} {\dot \phi}^{2}   \right)}^{.} \right] Q_{\phi} =0 \nonumber\\
\ddot \Psi_{r,lm} + 3 H \dot \Psi_{r,lm}+ \left( \frac{k^{2}}{a^{2}}
+
M^2_{r,l}(\phi) \right) \Psi_{r,lm} =0 \, . %
\ea%
 Here $k$ is the momentum
number in the Fourier space. One interesting aspect of these
equations is that the adiabatic and iso-curvature modes completely
decouple and they do not source each other. This is a result of our
initial conditions in which $\Psi_{r,lm}$ fields are turned off.
Because of this, the equation of motion for $\Psi_{r,lm}$ is that of
a scalar field in a homogeneous and isotropic expanding background
but with a time dependent mass.

These
equations are accompanied by Einstein equations
\ba
\label{Einstein}
3H ( \dot \Phi +
H \Phi) + \left(\dot H + \frac{k^{2}}{a^{2}} \right) \Phi& =&
-\frac{1}{2M_{P}^{2}} \left[ \dot \phi \dot {\delta \phi} + V_{0 \,
,\phi} \delta \phi
\right]  \nonumber\\
\dot \Phi + H \Phi &=& \frac{1}{2M_{P}^{2}}  \dot \phi \delta \phi
\, . \ea Interestingly enough,  $\Psi_{r,lm}$ fields do not show up
in perturbed Einstein equations due to our assumption that they are
absent at the background dynamics. Physically, this means that the
modes $\Psi_{r,lm}$ do not carry energy up to linear order in
perturbation theory. Since they do not couple to gravitational
potential and the inflaton perturbations, they have no gravitational
effect, justifying the name iso-curvature perturbations.

Consider the normalized curvature perturbation ${\cal R}$ and the
normalized iso-curvature perturbations ${\cal S}_{r,lm}$%
 \ba\label{RS}%
  {\cal R }
\equiv \frac{H}{\dot \phi} Q_{\phi} \quad , \quad {\cal S }_{r,lm}
\equiv \frac{H}{\dot \phi}  \Psi_{r,lm} \, . \ea %
Using the Einstein
equations one can show %
\ba\label{dot R}%
 \dot {\cal R} = \frac{H}{\dot H}
\frac{k^{2}}{a^{2}} \Phi \, .%
\ea%
 This indicates that on arbitrary
large scales where $k \rightarrow 0$, the curvature perturbation is
conserved. This is similar to single field inflationary system.

Here we assume that initially $\delta \phi$ and $\Psi_{r,lm} $ are
random, Gaussian and adiabatic  fields which are excited quantum
mechanically from vacuum and%
 \ba\label{vac}%
  \langle Q^{\star}_{\phi \,
{\bf k}}  \,  Q_{\phi\,  {\bf k'}}  \rangle & =& \frac{2
\pi^{2}}{k^{3}}
P_{Q_{\phi}} \delta^{3} ({\bf k} -{\bf k'} )  \nonumber\\
\langle {\Psi^\star_{r,lm\ {\bf k} } }  \, \Psi_{r',l'm'\ {\bf k'} }
\rangle &=&    \frac{2 \pi^{2}}{k^{3}}
P_{\Psi_{r,l}} \, \delta_{r r'}  \,\delta_{l l'}  \delta_{m m'} \,  \delta^{3} ({\bf k} -{\bf k'} )   \nonumber\\
\langle Q^{\star}_{\phi \,  {\bf k}} \Psi_{r,lm\ {\bf k'} }
\rangle&=& 0 \, , \ea %
where  $P_{Q_{\phi}} $ and $P_{\Psi_{r,l}}$ are the primordial power
spectra of the scalar fields. We note that the last equation above
also holds during the inflationary stages, indicating that there is
no cross-correlation between adiabatic and iso-curvature modes.
Physically, this means that they do not source each other as can be
seen from \eqref{KG}.

%%%%%%%%%%%%%%%%%%%%%%%%%%%%%%%%%%%%%%%%%%%%%%

\subsection{Power spectra from Hubble-crossing to end of
inflation}\label{power-spectrum-to-inf-end}%

As usual, it is instructive to express the Klein-Gordon equations
(\ref{KG}) in terms of variables $u$ and $v_{r,m} $ defined by
$$u\equiv  a Q_{\phi} \quad , \quad  v_{r,lm}  \equiv a \Psi_{r,lm} \, .$$
Going to conformal time $d \,  t =  a \, d\tau$, (\ref{KG}) is
transformed into Schr\"odinger equation form
\begin{subequations}%
\label{Schr}\begin{align}
\frac{d^{2} u}{d \tau^{2}} +  \left[ k^{2 } - \frac{2- 3 \eta + 9 \epsilon}{ \tau^{2}} \right] u&=0 \\
\frac{d^{2}  v_{r,lm}  }{d \tau^{2}} +  \left[ k^{2 } - \frac{2- 3
\eta_{r,l} + 3 \epsilon}{ \tau^{2}} \right]  v_{r,lm} &=0
\end{align}\end{subequations}%
 where  $\eta_{r,l} = M_{r,l}^{2}(\phi)/3 H^{2}$ given
by  (\ref{M-slow-roll}).

The initial conditions deep inside the Hubble radius are given by
the Bunch-Davies vacua. We evolve equations in (\ref{Schr}) till the
time of horizon crossing $\tau_{*}$ at which $k= (a H)_{*}$. Here
and below the subscript $*$ indicates that the quantities are
calculated at the time of Hubble crossing during inflation.
Following the standard inflationary prescriptions e.g.
\cite{Bassett:2005xm}, up to the first order in slow-roll parameters
one has \ba u&=& \frac{\sqrt{ \pi |\tau|} }{2} \, e^{  i (1+ 2
\nu_{R}) \pi/4  }  \, H^{(1)}_{\nu_{R}} ( k | \tau|)
\nonumber\\
v_{r,lm} &=&   \frac{\sqrt{ \pi |\tau|} }{2} \, e^{  i (1+ 2
\nu_{S_{r,lm}}) \pi/4  }  \, H^{(1)}_{\nu_{S_{r,lm}}} ( k | \tau|)%
\ea%
 where $H^{(1)}(x)$ is the Hankel function of the first kind and
$\nu_{R}$ and $\nu_{S_{r,lm}}$ are given by%
\ba%
\nu_{R} = \frac{3}{2} + 3 \epsilon - \eta \quad , \quad
\nu_{S_{r,lm}} = \frac{3}{2} + \epsilon  - \eta_{r,l} \, .%
\ea%
 Correspondingly, using (\ref{RS}), the power spectra of the
scalar curvature perturbations and iso-curvature perturbations,
$P_{{\cal R}}$  and $P_{ { \cal S }_{r,lm}  }$, at the time of
Hubble exit are given by \ba \label{PR} P_{{\cal R}}|_{\star}
&\simeq& \left( \frac{H^{2}}{2 \pi \dot \phi}  \right)_{\star}^{2}
\left[  1+ (-2 + 6 C) \epsilon - 2C \eta  \right]_{\star} \nonumber\\
P_{ { \cal S }_{r,lm}  }|_{\star} &\simeq& \left( \frac{H^{2}}{2 \pi
\dot \phi}  \right)_{\star}^{2} \left[  1+ (-2 + 2 C) \epsilon - 2C
\eta_{r,l}  \right]_{\star} \, . %
\ea%
 Here $C= \Gamma'(3/2)/\Gamma(3/2)+ \ln 2 \simeq 0.7296$  where $\Gamma(x)$ is the Gamma function.

Few e-folds after the mode of interest has left the Hubble radius, one can
neglect the term $k^{2}/a^{2}$  in  (\ref{KG}). Furthermore, it
would be more instructive to use the number of e-folds before the
end of inflation, $N_{e}$, as the clock. Using the relation $ d
N_{e} = H dt$, (\ref{KG}) is transformed into%
\ba\label{evol1}%
 \frac{d Q_{\phi}}{Q_{\phi}} \simeq  (2 \epsilon -
\eta)\,   d N_{e} \quad , \quad \frac{d \Psi_{r,lm} }{ \Psi_{r,lm}}
\simeq - \eta_{r, l}  \, d N_{e}  \, .
%\frac{M(\phi)^{2}}{3 H^{2}} d N_{e} \, .
\ea %
To calculate  the power spectra as a function of $N_{e}$, for
$\cal R$ and ${\cal S}_{r,lm}$ defined as in  (\ref{RS}), we need to
incorporate the evolution of the pre-factor $H/\dot \phi$ which is
given by \ba \label{evol2} \frac{d}{d N_{e}} \left( \frac{H}{\dot
\phi} \right) \simeq - (2 \epsilon - \eta) \left( \frac{H}{\dot
\phi}\right) \, . \ea Combining equations (\ref{evol1}) and
(\ref{evol2}) one obtains \ba \label{PRN}
P_{{\cal R}}(N_{e})&\simeq& P_{{\cal R}}|_{*}  \\
\label{Ps} P_{ {\cal S}_{r,lm} } (N_{e})&\simeq&  P_{ {\cal
S}_{r,lm} }|_{*}  \exp \left[ -2 \int_{0}^{N_{e}} d N_{e}'
B_{r,l}(N_{e}') \right]
\, , %
\ea%
 where $B_{r,l}(N_{e}) \equiv  2 \epsilon - \eta + \eta_{r,l}$ and
using  (\ref{M-slow-roll}) one obtains%
%\ba\label{B(N)}
$$ B_{r,l}(N_{e})\simeq  2 \epsilon +  \omega (2+
\omega) \eta
- sgn(V_{0}')  \frac{\sqrt{2\epsilon} M_{P}}{\phi}
(4\omega+3)(\omega+2)
+\frac{6M_{P}^2}{\phi^2}
(\omega+1)(\omega+2)  \, $$
%\ea
 where $\omega$ takes values $-1$, $-l-2$ and $l-1$ respectively for
 zero, $\alpha$ and $\beta$ modes (\emph{cf.} section
 \ref{mass-spectrum-section}).

 As explained previously, due to conservation of $\cal R$ on super-Hubble scales,
 $P_{{\cal R}}$ remains unchanged after Hubble exit as shown in  (\ref{PRN}).
 On the other hand, the evolution of power spectra for the iso-curvature modes,
 $ P_{ {\cal S}_{r,lm} } (N_{e})$,  as shown in (\ref{Ps}), depends  on
 the dynamics of the inflationary background and  the eigenvalues $\omega$.

%%%%%%%%%%%%%%%%%%%%%%%%%%%%%%%%%%%%%%%%%%%%%%
\subsection{Iso-curvature vs. entropy perturbations}
\label{iso}

As we saw in the previous section, the $\Psi_{r,lm}$ perturbations
do not couple to the inflaton field and the gravitational potential
so they do not contribute to the primordial curvature perturbations
during inflation which justify the name iso-curvature for these
modes. Physically, this means that up to the first order in
perturbation theory, $\Psi_{r,lm}$ fields do not carry energy during
inflation. As showed in section \ref{consistent-truncation-section}
this has the origin in our initial conditions for the $\Psi_{r,lm}$
fields that they are absent in classical background dynamics.

In the literature, the terminologies ``iso-curvature perturbations''
and  ``entropy perturbations'' are usually used interchangeably.
However, one can easily check that $\Psi_{r,lm}$ perturbations do
not induce entropy perturbations during inflation. To see this
explicitly, let us look at non-adiabatic components of pressure,
$\delta p_{nad}$, defined as \cite{Wands:2007bd} \ba \delta p_{nad}
= \delta p - \frac{\dot p}{\dot \rho} \delta \rho \, , \ea where
$\rho$ and $p$ are the background energy density and pressure
respectively and $\delta \rho$ and $\delta p$ represent their first
order variations. Using Einstein equations (\ref{Einstein}) one can
show that \ba \delta p_{nad} \simeq - 4 M_{P}^{2}
\frac{k^{2}}{a^{2}} \Phi \,  , \ea which is the same as the standard
single-field inflation result. On super-Hubble scale one observes
that $\delta p_{nad}$ vanishes and there is no entropy perturbation.
This is similar to our earlier result  (\ref{dot R}) that $\dot
{\cal R} \simeq 0$ on super-Hubble scale. In this work, in order to
emphasis that there is no entropy perturbations in our setup, we use
the terminology iso-curvature perturbations throughout.

In general multiple-field inflation, the iso-curvature perturbations
perpendicular to the classical inflationary trajectory do produce
entropy perturbations and $\delta p_{nad}\neq 0$. These entropy
perturbations source the adiabatic perturbation and have non-zero energy at the
linear order of perturbations so they also source the Einstein
equations and contribute to the primordial curvature perturbations.
More specifically \ba
 \dot{\mathcal{R}} = \frac{H}{\dot H}\frac{k^2}{a^2} \Phi +
 2\sum_{r,\ lm}\dot{\theta}_{r,lm}\mathcal{S}_{r,lm} \ .
 \ea%
The above equation is a generalization of the two-field  result of
\cite{Gordon:2000hv},  the sum in the last term is over $3N^2-1$
entropy modes and ${\theta}_{r,lm}$ represent the angle between
$\Psi_{r,lm}$ and the inflaton trajectory, $\phi$ in our case, in
the space of $3N^2$ scalar fields. As discussed above, if we start
with the initial conditions $\Psi_i=0,\ \dot\Psi_i=0$, $\phi$
behaves as the sole inflaton field and in the language of
\cite{Gordon:2000hv}, the background inflationary trajectory in
$\phi-\Psi_{i}$ phase space is flat, i.e. $\dot\theta_{r,lm}=0$.
 Therefore, these iso-curvature perturbations do not feed the
curvature perturbation and there is no cross-correlation between the
iso-curvature and adiabatic perturbations.

However, if we start with an arbitrary initial condition in
$\phi-\Psi_{i}$ field space, $\Psi_{i}$ are not frozen classically, the
inflation trajectory is curved and the inflaton field has a
component along the $\Psi_{r,lm}$ direction and $ \dot\theta_{r,lm}
\neq 0$. This in turn produces entropy perturbations during
inflation which also source the cosmic perturbations.

Having this said, however, there are two mechanisms to create
entropy perturbations in our setup. The first way is to consider
second order perturbation in $\Psi_{r,lm}$. In second order
perturbation theory, $\Psi_{r,lm}$ carries energy
\cite{Durrer:1994nk} and  couple to both Einstein equations and the
inflaton field perturbation $\delta \phi$. This in turn leads to entropy perturbations
during inflation. However, the amplitude of these entropy
perturbations are much smaller than the first order adiabatic
perturbations coming from $\delta \phi$. The second mechanism to
create entropy perturbations from iso-curvature perturbations
$\Psi_{r,lm}$ can happen during preheating and/or reheating era.
Similar idea was used in \cite{Byrnes:2005th} through asymmetric
preheating.

%%%%%%%%%%%%%%%%%%%%%%%%%%%%%%%%%%%%%%%%%%%%%%%%%%%%%%%%%%%%%%%%%%%%%%%%%%%%%%%%%%%%%%%%%%%

\subsection{  Curvature and iso-curvature perturbations power spectra for specific examples   }

Using the general formulation presented in previous subsections we
calculate the power spectra at the end of inflation for chaotic,
symmetry breaking and the inflection point inflationary potentials.

%%%%%%%%%%%%%%%%%%%%%%%%%%%%%%%%%%%%%%%%%%%%%%

 \subsubsection{Chaotic inflation $\frac{m^{2}}{2} \phi^{2}$}

 When $\lambda= \kappa=0$, the action takes the form%
 \ba
 \label{chaotic1}%
S= -\frac{1}{2} \partial_{\mu} \phi  \partial^{\mu} \phi -
\frac{1}{2} \partial_{\mu} \Psi^\star_{r,lm}  \partial^{\mu}
\Psi_{r,lm} - \frac{1}{2} m^{2} \left[ \phi^{2} + \Psi^\star_{r,lm}
\Psi_{r,lm} \right]  \, .
 \ea
The $3N^2$ modes all have equal masses. Interestingly, this closely
resembles the chaotic inflation studied in
\cite{Brandenberger:2003zk, Byrnes:2005th}. The potential
(\ref{chaotic1}) has $O(3N^{2})$ symmetry for the fields $\phi$ and
$\Psi_{r,lm}$.  The background inflation field is $\phi$ and the
number of e-folds in terms of   the initial value of the inflaton
field $\phi_{i}$ is given by
 \ba
 N_{e} \simeq  \frac{ \phi_{i}^{2}}{4 M_{P}^{2}} \, .
 \ea
 Correspondingly, the amplitude of the adiabatic curvature perturbation  is
 \ba
 P_{{\cal R}} \simeq \frac{1}{6 \pi^{2}} \left(\frac{m N_{e}}{M_{P}} \right)^{2} \, .
 \ea

  %%%%%%%%%%%%%%%%%%%%%%%%%%%%%%%%%%%%%%%%%%%%%%
\begin{figure}[t]
\includegraphics[angle=0, width=74mm, height=70mm]{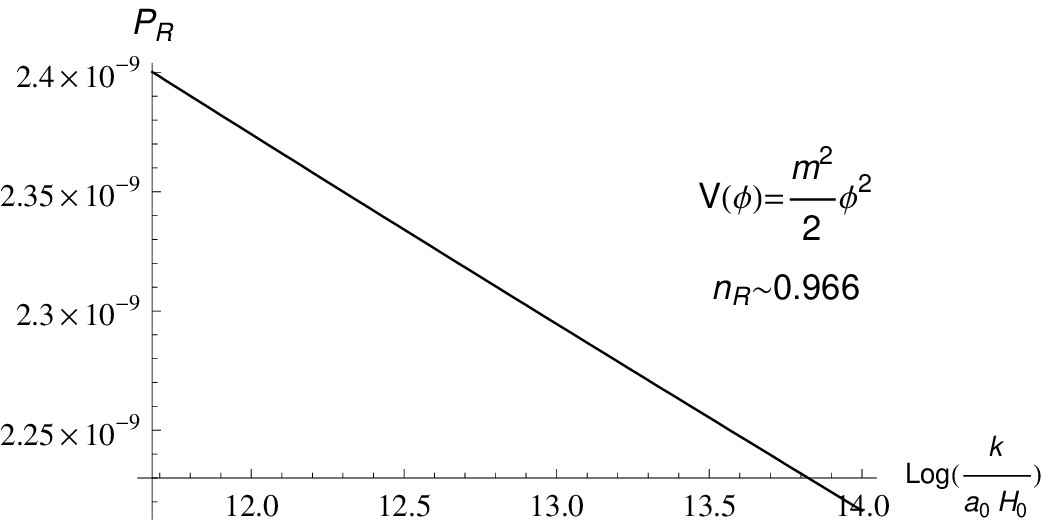}
\includegraphics[angle=0, width=74mm, height=70mm]{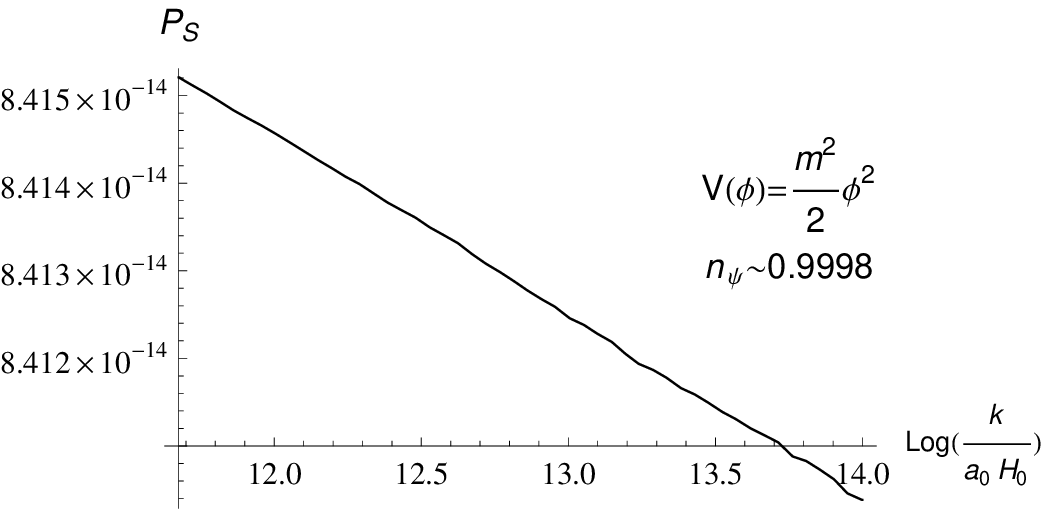}
\vspace{0.5cm} \caption{Left and right graphs respectively show the
curvature and isocurvature spectra for chaotic inflation with
potential $\frac{m^2}{2}\phi^2$}\label{m^2phi^2}
\end{figure}
%\vspace{1cm}
%\vspace{1cm}
%%%%%%%%%%%%%%%%%%%%%%%%%%%%%%%%%%%%%%%%%%%%%%

\begin{figure}[t]
\includegraphics[angle=0, width=74mm, height=70mm]{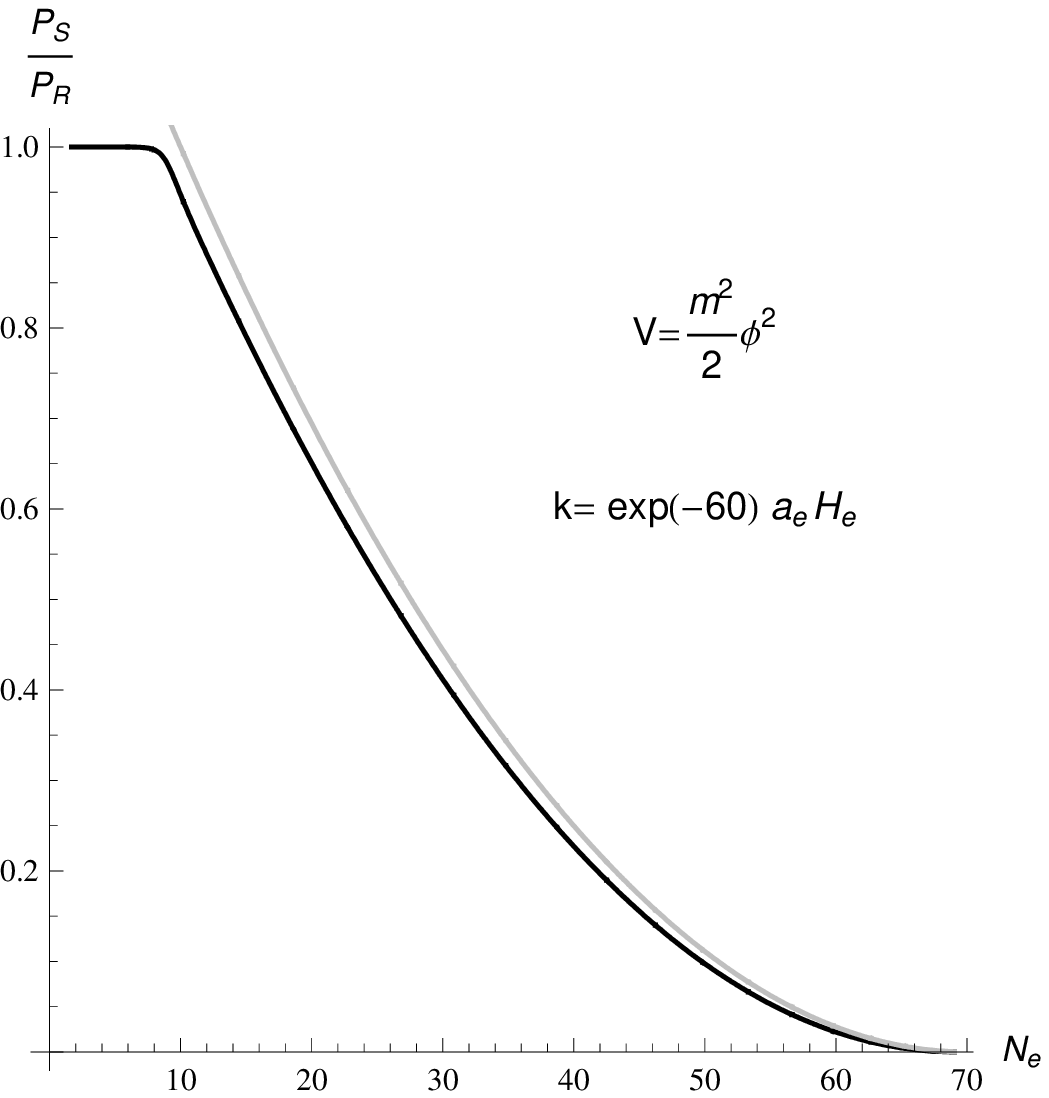}
\includegraphics[angle=0, width=74mm, height=70mm]{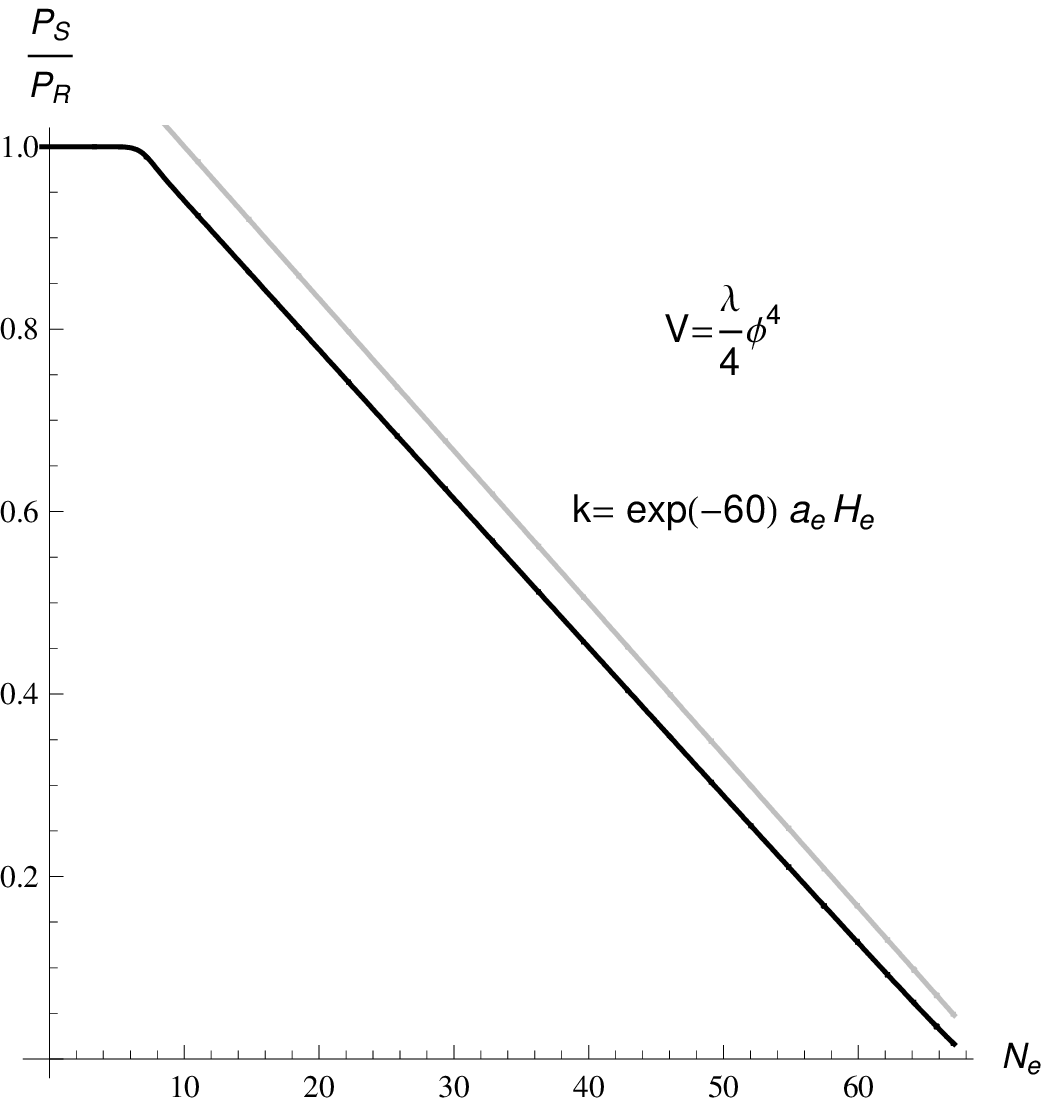}
\vspace{1cm} \caption{Left graph shows the ratio $P_{ {\cal
S}_{r,lm} } (N_{e}) /P_{{\cal R}}|_{*}$ vs. $N_e$ for the mode that
exit the Hubble radius 60 e-folds before the end of chaotic
inflation with $V=\frac{1}{2}m^2 \phi^2$. Right graph shows the
evolution of $P_{ {\cal S}_{\beta,3m} } (N_{e}) /P_{{\cal R}}|_{*}$
for $l=1,2$ $\beta-$mode in $\lambda_{eff} \,  \phi^4/4$ potential.
Black and gray curves respectively demonstrate the numerical and
analytic results.}\label{PS/PR-chaotic}
\end{figure}

For $N_{e}=60$, to fit the WMAP5 normalization $P_{{\cal R}} \simeq
2.41 \times 10^{-9}$, one
 requires $m \simeq 6.304 \times 10^{-6} M_{P}$. For such values of
 $m$, we have calculated numerically the amplitudes of $3N^2-1$ iso-curvature spectra at the end of inflation, \textit{i.e.} where $\epsilon=1$, which turns out to be $P_{ {\cal S}_{r,lm} } \simeq 8.426 \times 10^{-14}$ at today's Hubble scale. The spectral indices of adiabatic and
 iso-curvature spectra at such scale, respectively, are $n_{\cal R} \simeq 0.966$ and
$n_{ \Psi_{r,lm} }\simeq 0.9998$, which coincide with the analytic results of
 \cite{Byrnes:2005th}, see {\bf Fig.\ref{m^2phi^2}}. One can lower the value of  $m$ -- which in
 turn lowers the amplitude of adiabatic perturbations -- and generate
 the difference by transforming the iso-curvature fluctuations to curvature ones through an asymmetric
 mechanism of preheating \cite{Byrnes:2005th}. The ratio $P_{ {\cal S}_{r,lm} }/P_{{\cal R}}$
 for the mode that exit the horizon $60$ e-folds before the end of
inflation is graphed as a function of $N_e$, see {\bf
Fig.\ref{PS/PR-chaotic}}. Note that as in this case all the
$\Psi_{r,lm}$ modes have the same mass and hence $\eta_{r,l}$, $P_{
{\cal S}_{r,lm} }/P_{{\cal R}}$ are independent of $r$ and $l,m$.
 When the mode is inside the Hubble radius the spectra are almost equal and the ratio is one. However, around the Hubble-exit, the  iso-curvature mode starts to decay following
 (\ref{Ps}). For the mode that exits the Hubble radius $60$ e-folds
 before the end of inflation, $k_{60}= e^{-60} a_e H_e$,  according to (\ref{Ps})
the ratio decays like
 \ba%
 \frac{P_{ {\cal S}_{r,lm} } (N_{e}) }{P_{{\cal R}}|_{*}}
 \simeq {(1-N_e/60)}^2 \, ,
 \ea%
where $N_e$ is the number of e-folds the mode spends outside the
Hubble radius before the end of inflation. As  can be seen in {\bf
Fig.\ref{PS/PR-chaotic}}, the analytic result is in a good agreement
with the numerical
 one.

 %%%%%%%%%%%%%%%%%%%%%%%%%%%%%%%%%%%%%%%%%%%%%%
%\vspace{1cm}

%%%%%%%%%%%%%%%%%%%%%%%%%%%%%%%%%%%%%%%%%%%%%%

 For this model, the amplitude of tensor fluctuations of the adiabatic mode at today's Hubble scale is $P_{T}(k_{60})\simeq 3.1856 \times
 10^{-10}$. The corresponding tensor/scalar ratio, $r$, is $ r \simeq 0.132$.
The tensor spectral index, $n_T \equiv d\ln P_T/d\ln k$, is  $ n_{T} \simeq
 -0.0165$. Future CMB probes such as PLANCK \cite{PLANCK},
 or exclusive polarization probes such as CMBPOL \cite{Baumann:2008aq}  or QUIET \cite{quiet} should be able to test this scenario.

%%%%%%%%%%%%%%%%%%%%%%%%%%%%%%%%%%%%%%%%%%%%%%
\subsubsection{Chaotic inflation $\frac{\lambda_{eff}}{4} \phi^{4}$}

 When $m = \kappa=0$, the potential energy $V= V_{0} (\phi) + V_{(2)}(\phi, \Psi_{i})$  is
 \ba
 V= \frac{\lambda_{eff}}{4} \phi^{4}
 + \frac{\lambda_{eff}}{4}  (\omega^{2} - \omega) \,   \phi^{2}\,\Psi^\star_{r,lm}\Psi_{r,lm} \, .
 \ea
To match the amplitude of adiabatic perturbations with WMAP result
at horizon scale, $\lambda_{eff}\simeq 1.6315 \times 10^{-13}$. For
such value of $\lambda_{eff}$, the scalar spectral index for the
adiabatic mode is $n_{\cal R} \simeq 0.949$.

In $\lambda_{eff} \phi^4/4$ chaotic model, the masses of
iso-curvature modes are different. The lowest mass belongs to the
$l=1,2$  $\beta-$modes whose mass is equal to zero,
$M^2_{\beta,1}(\phi)=0$. The corresponding iso-curvature spectrum
amplitude at the end of inflation is equal to $P_{ {\cal
S}_{\beta,1m} }\simeq3.949 \times 10^{-11}$ at today's Hubble scale.
The corresponding spectral index is $n_{ \Psi_{\beta,1m} } \simeq
0.966$. The relatively larger value of iso-curvature perturbations
could be attributed to the fact that the iso-curvature spectrum for
this mode decays linearly with the number of e-folds it spends
outside the horizon (see below).

The next modes in the tower of masses are the zero modes and $l=3$
$\beta-$modes whose masses are equal to $M^2_{\beta,3}(\phi)=
\lambda_{eff} \phi^2$. Their corresponding amplitudes are $   P_{
{\cal S}_{\beta,3m} }\  \simeq 4.449 \times 10^{-13}$ at today's
horizon scale. Taking the multiplicity of $\alpha$ and $\beta$ modes
into account, there are $N^2+6$ iso-curvature modes with such an
amplitude. The corresponding spectral index is $n_{ \Psi_{\beta,3m}
} \simeq 0.9828$.

For the $l=0$ $\alpha-$mode with the mass  $M^2_{\beta,4}(\phi)=3
\lambda_{eff} \phi^2$, which is equal to the mass of the $l=4$
$\beta-$mode, the iso-curvature spectrum amplitude is $  P_{ {\cal
S}_{\beta,4 m} }   \simeq  3.967\times 10^{-18}$ at today's Hubble
scale.  The corresponding spectral index is equal to $  n_{
\Psi_{\beta,4 m} }       \simeq 1.016$, which indicates a blue
spectrum.

In general mass of a $l\geq 0$ $\alpha-$mode is identical to the
$l+4$ $\beta-$mode. Therefore, there are $4l+9$ iso-curvature modes
with identical spectra, all of which have a blue tilt. Increasing
the value of $l$ for $\alpha-$mode and $\beta-$mode, the amplitude
of heavy iso-curvature modes decreases quickly. This can be
understood from (\ref{Ps}), which yield%
\ba%
\label{PS/PR}
 \frac{P_{ {\cal S}_{r,lm} } (N_{e}) }{P_{{\cal R}}|_{*}}
\simeq {(1-N_{e}/60)}^{ 1+\frac{\omega^{2} - \omega }{2} } =
\begin{cases}
{(1-N_{e}/60)}^{2} \quad \quad \quad  \quad \quad \quad \mathrm{zero \, \, modes}\\
{(1-N_{e}/60)}^{(l^{2} + 5 l  + 8)/2} \, \, \quad  \quad \alpha - \mathrm{modes}\\
{(1-N_{e}/60)}^{(l^{2} - 3l  + 4)/2} \, \,  \quad  \quad  \beta -
\mathrm{modes},
\end{cases}
\ea%
 where again $N_e$ is the number of e-folds the mode spends outside
the Hubble radius before inflation ends. In particular, for $l=1,2
\, \, \beta$-mode our analytic result indicates that the ratio $
P_{ {\cal S}_{\beta,3 m} } (N_{e}) /P_{{\cal R}}|_{*}    $ decreases
linearly with $N_{e}$, which is verified numerically as shown in
{\bf Fig. \ref{PS/PR-chaotic}}.

%We also note that as the mass increases, the spectral indices of the entropy modes increases.

Besides a considerable iso-curvature/adiabatic ratio of $1.638 \%$
for $l=1,2$  $\beta-$modes, another signature of the model is its
observable gravity waves. For this model the amplitude of tensor
spectrum for adiabatic perturbations at  Hubble scale today is
$P_{T}(k_0) \simeq 6.3176 \times 10^{-10}$, i.e. $r\simeq 0.26$,
whose spectral index is $ n_{T} \simeq -0.033$.  This model is
currently on the verge of becoming ruled out.

%%%%%%%%%%%%%%%%%%%%%%%%%%%%%%%%%%%%%%%%%%%%%%
\subsubsection{Symmetry breaking inflation}

We consider $\phi>\mu$  and $\mu/2< \phi<\mu$ cases separately. \;\
 \bigskip

 { \bf (a)  }  $\phi>\mu$

 \bigskip
 In this case neither of the parameters of the potential,
$\lambda$, $\kappa$ and $m$ are zero. To match the observational
constraints from WMAP5, the above parameters have to take the
following values given in \eqref{phi>mu}:
\begin{equation}
% \nonumber to remove numbering (before each equation)
   \lambda_{eff}\simeq 4.91\times 10^{-14}, \quad m\simeq 4.074\times 10^{-6} M_{P}, \quad \kappa_{eff}\simeq 9.574\times 10^{-13}M_{P}.
\nonumber\end{equation} The scalar spectral index for the adiabatic
perturbations is $ n_{\cal R} \simeq 0.959$.

The lowest masses in the tower of $\Psi_{r,lm}$ iso-curvature modes
belong to the zero mode whose mass is $M^2_{0}(\phi)= \lambda_{eff}
\phi^2-2\kappa_{eff} \phi+m^2$. The amplitudes and spectral indices
of these two iso-curvature spectra are respectively $ P_{ {\cal
S}_{0} }   \simeq 1.162 \times 10^{-11}$ and $ n_{ \Psi_{0 } }
\simeq 0.981$.

Next in the tower of iso-curvature modes is the $l=1$ $\beta-$mode
whose mass is equal to the inflaton mass, $M^2_{\beta,1}(\phi)= m^2$. Its amplitude and index are,
respectively, $  P_{ {\cal S}_{\beta,1} }   \simeq 1.131\times
10^{-12}$ and $   n_{ \Psi_{\beta,1} } \simeq 0.978$. The $l=2$ $\beta-$mode with mass equal to
$2\kappa_{eff} \phi+m^2$ stands next in the tower. The amplitude of
this mode is equal to $ P_{ {\cal S}_{\beta,2 m} } \simeq 8.842
\times 10^{-18}$. The corresponding iso-curvature spectrum for this
mode has a blue tilt but an almost scale-invariant spectrum, with $
n_{ \Psi_{\beta,2} }      \simeq 1.002$. As before, the next
iso-curvature modes have negligible amplitudes at Hubble  scale and
therefore their contributions could be ignored.

The amplitude of tensor spectrum at Hubble scale is $P_{T}(k_{60})
\simeq 4.84\times 10^{-10}$, $r\simeq 0.2$ with the spectral index $
n_{T} \simeq -0.025$. Planck \cite{PLANCK} should be able to verify
this model.

%%%%%%%%%%%%%%%%%%%%%%%%%%%%%%%%%%%%%%%%%%%%%%
 \bigskip

 { \bf (b)  }  $\mu/2< \phi<\mu$
\bigskip

Here to satisfy the constraints from the amplitude and spectral
index from WMAP5, one has to adjust the the parameters as in
\eqref{phi<mu}:
\begin{equation}
% \nonumber to remove numbering (before each equation)
   \lambda_{eff}\simeq 7.187\times 10^{-14}, \quad m\simeq 6.824\times 10^{-6} M_{P}, \quad \kappa_{eff} \simeq 1.940\times 10^{-12} M_{P}.
\nonumber\end{equation} The index of the adiabatic spectrum for such
values of parameters is $n_{\cal R} \simeq  0.961$.

Again the least massive iso-curvature modes are the zero modes.
Their amplitude and spectral index are, respectively, $   P_{ {\cal
S}_{0} }   \simeq   1.46\times 10^{-11}$ and $    n_{ \Psi_{0 } }
\simeq    0.987$. The next biggest iso-curvature amplitude belongs
to the $l=1$ $\beta-$mode whose amplitude and spectral index are $
P_{ {\cal S}_{\beta,1} }   \simeq 6.55\times 10^{-16}$ and $ n_{
\Psi_{\beta,1} } \simeq 1.0545$, respectively. The $l=1$
$\alpha-$mode stands in the next rank with $
P_{ {\cal S}_{\alpha,1} }   \simeq 4.69\times 10^{-19}$ and $ n_{
\Psi_{\alpha,1}} \simeq 1.007$. Next iso-curvature modes have
smaller amplitudes that are completely negligible in comparison with the adiabatic one.

The amplitude of tensor spectrum at Hubble scale is $  P_{T}(k_0)
\simeq  1.307\times 10^{-11}$, i.e. $r\simeq 0.048$ with the
spectral index $ n_{T}   \simeq  -0.006$. Such gravity wave spectrum
could be detected by CMBPOL \cite{Baumann:2008aq} or QUIET
\cite{quiet}. The tensor spectrum is very close to being
scale-invariant in this case.

%%%%%%%%%%%%%%%%%%%%%%%%%%%%%%%%%%%%%%%%%%%%%%
 \bigskip

 { \bf (c)  }  $0< \phi<\mu/2$
\bigskip

The involved parameters have to be set at the values mentioned for region ({\bf b}) to satisfy the CMB constraints on the amplitude and scalar spectral index of adiabatic perturbations. However as the mass expressions for zero, $\alpha$ and $\beta$ modes, do not respect the symmetry $\phi \rightarrow -\phi + \mu$, the predictions for the isocurvature modes are in general different. In particular the largest isocurvature spectrum is produced by $l=1$ $\alpha-$mode whose amplitude and spectral index at the current Hubble scale would be $P_{ {\cal S}_{\alpha,1} }   \simeq 1.213\times 10^{-11}$ and $ n_{\Psi_{\alpha,1} } \simeq 0.953$. The zero modes stand next in the tower with the amplitude  of $3.84\times 10^{-14}$ whose spectrum is almost scale-invariant, $n_{\Psi_{0} } \simeq 1.006$. The next one is $l=1$ $\beta-$mode, whose mass is the same as the corresponding mode in the region ({\bf b}) and therefore its amplitude and spectral index are the same. The amplitudes of other 
 isocurvature modes are more suppressed.

%%%%%%%%%%%%%%%%%%%%%%%%%%%%%%%%%%%%%%%%%%%%%%
\subsubsection{Inflection point inflation}

A possible parameter set that satisfies the observational
constraints are:
\begin{equation}
% \nonumber to remove numbering (before each equation)
   \lambda_{eff}\simeq 4.8\times 10^{-14}, \quad m\simeq 10^{-6}M_{P}, \quad \kappa_{eff}\simeq1.94\times 10^{-12}M_{P}.
\end{equation}
For this parameter set the  scalar spectral index is $n_{\cal R} \simeq 0.93$ which is within
$2\sigma$ error bar of WMAP5 but is somewhat to its lower end.

The largest amplitude for isocurvature perturbations is obtained for $l=0$ $\alpha-$mode whose mass is $ M^2_{\alpha,0}(\phi)=
\lambda_{eff} \phi^2-2\kappa_{eff}\phi+m^2$. Its
corresponding spectrum amplitude and spectral index are $   P_{
{\cal S}_{\alpha,0} }   \simeq 1.755\times 10^{-16}$ and $   n_{
\Psi_{\alpha,0} } \simeq 1.00125$, at Hubble scales. The next modes in the
series is the $l=1$ $\alpha$-mode with
$M^2_{\alpha,1}=6 \lambda \phi^2-6\kappa \phi+m^2$ whose amplitude and spectral index are respectively $ P_{ {\cal S}_{\alpha,1} } \simeq
2.93\times 10^{-32}$ and $ n_{ \Psi_{0} } \simeq 2.17$. Other
iso-curvature modes are more suppressed in comparison
with the adiabatic one.

The amplitude of the tensor spectral index is negligible for the
above set of parameters in this model, $P_T\simeq 1.168 \times
10^{-13}$, which is scale-invariant with the precision of $10^{-6}$.

%%%%%%%%%%%%%%%%%%%%%%%%%%%%%%%%%%%%%%%%%%%%%%%%%%%%%%%%%%%%%%%%%%%%%%
\section{End of inflation and preheating}\label{preheating-section}

As  discussed while the $\phi$ field is turned on during inflation,
the other fields $\Psi_{r,lm}$ are also present. Although not turned
on at the onset of inflation (by the choice of initial conditions)
and hence due to the specifics of the classical dynamics of our
model remain zero during inflation, the presence of $\Psi_{r,lm}$
fields can be felt through quantum effects. These quantum effects
show up in two different contexts; one is of course through the
power spectrum of the quantum fluctuations of these fields at the
super-horizon scales, the iso-curvature modes which were discussed
in some detail in the previous section. The other quantum effect is
the possibility of  creation of the $\Psi_{r,lm}$ particles, due to
the coupling to the inflaton field $\phi$. If the pair creation
mechanism is ``efficient enough'' this will eventually back react on
the classical dynamics of the system. This effect, if too efficient
during inflation and before completion of the needed 60 e-folds, can
tamper the whole M-flation scenario.\footnote{This ``slow-down''
effect via particle creation, although potentially harmful for the
standard slow-roll inflation, can be used as the mechanism to render
an otherwise fast-roll inflationary scenario which does not give
enough e-folds, to an effectively slow-roll inflation with enough
number of e-folds. The recent publication \cite{trapped-inflation}
discusses this possibility.} Recalling the large number of
$\Psi_{r,lm}$ modes ($3N^2-1$), their collective effect on the
inflaton field could be very large ending inflation too fast. While
if activated only toward the end of inflation it will be a positive
feature of our model, providing us with a mechanism to end inflation
while transferring the potential energy of the inflaton field into
the kinetic energy of the $\Psi_{r,lm}$ fields, a preheating
scenario \cite{preheating-I, preheating-II, Bassett:2005xm,
preheat-review}.

In the first subsection, we first show that particle creation
during slow-roll inflation is not harmful to our M-flation model. In
the next subsection  we explore the possibility of the particle
creation as the basis for a natural and inherent preheating scenario
in our M-flation model.

%%%%%%%%%%%%%%%%%%%%%%%%%%%%%%%%%%%%%%%%%%%%%
\subsection{Particle creation during slow-roll inflation  is not harmful to
M-flation}\label{not-harmful-section}

In this section we show that quantum production of $\Psi_{r,lm}$
modes during inflation is not large enough to derail the slow-roll
M-flation. This is done at two steps, first we compute the particle
creation rate during inflation and then study the back reaction of
the $\Psi_{r,lm}$ modes on the dynamics of the inflaton $\phi$.

%%%%%%%%%%%%%%%%%%%%%%%%%%%%%%%%%%%%%%%%%%%%%%
\subsubsection{Quantum production of $\Psi_{r,lm}$ modes during inflation}
\label{infpsi}

The Lagrangian governing the dynamics of our model, up to the second
order in $\Psi_{r,lm}$ fields, and the corresponding equations of
motion are given respectively by  (\ref{S2}) and (\ref{KG}). As
explained before, the time-dependence in $M_{r,l}(\phi)$ leads to
$\Psi_{r,lm}$ particle creation from the vacuum
\cite{Traschen:1990sw, preheating-I, preheating-II, Bassett:2005xm,
preheat-review}. (Note that the  expression (\ref{massM}) for
$M_{r,l}^2 (\phi)$ depends on $r$ and $l$ as well as the momentum
number $k$.)

By replacing%
\be%
\chi_{r,lm\ k}=a^{3/2}\ \Psi_{r,lm\ k}\ , %
\ee%
the equation of  $\Psi_{r,lm}$ field takes the form of
\be\label{chi-e.o.m-general}%
\ddot\chi_k +\Omega_{k,rl}(t)^2\ \chi_k=0 \, ,
\ee%
 which is an oscillator with a time  dependent frequency, $\Omega^2_{k, rl}$%
\be\label{frequency}%
\Omega^2_{k, rl}=\frac{k^2}{a^2}-\frac94 H^2(1-\frac23\epsilon) + M_{r,l}^2(\phi)\ , %
\ee%
where  $\epsilon\equiv -\dot H/H^2$  which in the slow-roll limit
reduces to its conventional form \eqref{epsilon-eta}.

During inflation and when $\Omega^2<0$ and $\chi$ is either the
inflaton or iso-curvature perturbations, $\chi$ modes have an
imaginary frequency. As is well-known $\Omega^2<0$ during inflation
(e.g. see discussions and analysis of section
\ref{mass-spectrum-section}) happens for the ``super-horizon''
modes. These imaginary frequency modes are those which follow a
classical dynamics and contribute to the power spectrum of curvature
or iso-curvature (entropy) modes.

For  particle creation inside the horizon which is what we will be
mainly concerned with here and can happen if the time variation of
$\Omega^2$ is not negligible, we should focus on the $\Omega^2>0$
regime. Let us suppose that we have the solution to
\eqref{chi-e.o.m-general} for the $\Omega^2>0$ regime. Our goal is
to compute the number density of the $\chi_k$ particles produced
during inflation and for that we need to compare the solutions at
$t=-\infty$ (the onset of inflation) to $t=+\infty$ (the end of
inflation). Denoting the former by $\chi^-_k$ and the
latter by $\chi_k^+$, one may expand%
\be%
\chi_k^-=A_k\chi_k^+ + B_k\chi^{\ast}_k{}^+\ .%
 \ee%
The $A_{k}$ and $B_{k}$ coefficients maybe thought as the Bogoliubov
transformation parameters. The number density of $\chi_k$ mode
produced is then equal to $|B_k|^2$. For a detailed discussion on
this matter see \cite{preheating-I} or Appendix B of
\cite{beauty-attractive}.

Except for some specific cases
(e.g. see \cite{Greene-Kofman, Bassett:2005xm}), however, it is not
possible to solve \eqref{chi-e.o.m-general} and we are hence forced
to use approximations. One of the approximations which is usually
employed (e.g. see \cite{Bassett:2005xm, preheat-review,
preheating-II}) is the stationary phase approximation, which if
\eqref{chi-e.o.m-general} viewed as the Schrodinger equation, it is
the WKB approximation. This approximation is valid if%
 \ba%
  \dot
\Omega_{k, rl} \lesssim \Omega^2_{k, rl} \quad , \quad \Omega^2_{k,
rl}
>0 \, .%
 \ea%
In this regime one may expand $\Omega^2$ around its minima as%
\be\label{Omega-expand}%
\Omega^2_{k, rl}=\Omega^2_{0k, rl}+\Gamma^2_{k, rl} (t-t_k)^2+ {\cal
O}((t-t_k)^3),\quad\qquad \dot\Omega_{k, rl}|_{t=t_k}=0\ .%
\ee%
where $t_k$ is where $\dot\Omega_{k,rl}$ vanishes. Note that within
our assumptions $\Omega_{0k, rl}^2$ and $\Gamma^2_{k,rl}$ are both
positive. (If $\Omega^2_{k,rl}$ has several minima one should sum
over all of them.)

In the WKB approximation the number density of the particles produced for
each mode $k$ is%
\be\label{chi-produced}%
\langle\chi_k|\chi_k\rangle=|B_k|^2=e^{-\pi\frac{\Omega_{0,k}^2}{\Gamma_k}}\ .%
\ee%
As we see the particle creation is effective if
$\frac{\Omega_{0,k}^2}{\Gamma_k}$ is of order one and not large.

In order  to be specific and to
get a better theoretical understanding of the analysis we shall
focus on the specific $\lambda_{eff}\phi^4/4$ model for which
$\kappa_{eff}$
and $m^2$ both vanish. In this case%
\be\label{M2-phi4}%
M^2=\nu \lambda_{eff}\phi^2\ , %
\ee%
where $\nu=1, \frac12 (l+2)(l+3),\ \frac12 (l-1)(l-2)$ respectively
for
zero, $\alpha$ and $\beta$ modes. Moreover,%
\[%
\epsilon=\frac23\eta= 8\left(\frac{M_{P}}{\phi}\right)^2 .%
\]%
In the leading order in $\epsilon,\ \eta$%
\be\label{Omega-Gamma-phi4}%
\begin{split}%
\frac{k^2}{a^2H^2} &= \frac{1}{2H^4}\Gamma^2_{k, rl}=\frac34 \epsilon(3-\nu\epsilon)\ ,\cr%
\Omega_{0k, rl}^2 &=\frac32 H^2(\nu\epsilon-\frac32)\ .%
\end{split}%
\ee%
The creation of particles can then happen in the window%
\be\label{nu-window}%
\frac32 <\nu\epsilon <3\ .%
\ee%
That is, it is not possible for zero modes and for $\alpha$ and
$\beta$ modes with $l$ less or of order $1/\sqrt{\epsilon}$.
Recalling the expression \eqref{chi-produced} and the exponential
suppression by the factor of
$\Omega^2_0/\Gamma$, the particle creation is effective in the region%
\be\label{non-adiabatic}%
\frac{\Omega^2_{0k, rl}}{\Gamma_{k,
rl}}=\frac{\nu\epsilon-\frac32}{\sqrt{2\epsilon(1-\frac{\nu\epsilon}{3})}}
\lesssim 1\ . %
\ee%
The above can be satisfied if $\nu$ is close to its lower bound
$\frac{3}{2\epsilon}$. However, recalling that $\nu$ is
integer-valued and that for large $\nu$, $\nu\simeq l^2/2$, the
number of allowed $l$'s is very limited and as a result
$\nu\epsilon$ can be tuned around the center value $3/2$ with
accuracy of order $\sqrt{\epsilon}$.
That is, for $\nu$ around%
\be\label{delta-def}%
\nu\epsilon=\frac32+\delta \sqrt{\epsilon}\ ,%
\ee%
with $\delta$ being an order one number, the particle creation is
effective.

 As the last step we compute the energy which is carried by the  $\Psi_{r,lm}$
particles in $\alpha$ and $\beta$ modes produced during inflation.
For that we need to integrate over the number density
\eqref{chi-produced}, explicitly%
\be\label{Npsi}%
N_{\Psi}= \sum_{r,l} D_{r,l} \int\ \frac{d^3k}{(2\pi)^3}\
\frac{1}{a^3\Omega_{0k, rl}} \
e^{-\pi\frac{\Omega^2_{0k, rl}}{\Gamma_{k, rl}}}\ , %
\ee%
where sum over $r$ only runs over $\alpha$ and $\beta$ modes and
$D_{r,l}$ is the degeneracy of the modes which is equal to $2l+1$
for $\alpha$ and $\beta$ modes. Recalling that the particle
creation is effective in the range,%
\[
\frac{k}{a}=H\sqrt{\frac98\epsilon},\qquad \frac{k\Delta k}{a^2}=
\frac38 H^2 \epsilon^{3/2}\ \delta\, \qquad
\Omega_0=H\sqrt{\frac32\delta}\ \epsilon^{1/4} ,
\]%
and for $\nu\epsilon=\frac32$, $D_{r,l}$ is roughly
$2\sqrt{3/\epsilon}$. Given the above
one can perform the integral %
\be\label{Npsi-result}%
N_\Psi\simeq \frac{9}{8\pi^2}\ H^2\ \epsilon^{5/4}\
\delta^{1/2}\cdot
e^{-\pi\delta}\ . %
\ee%
We would like to remark that as $\Omega^2_0/\Gamma$ is of order
unity one may still trust the WKB approximation.

%%%%%%%%%%%%%%%%%%%%%%%%%%%%%%%%%%%%%%%%%%%%
\subsubsection{Back reaction on the inflationary dynamics}%

The created $\Psi$ particles during slow-roll inflation will back
react on the dynamics of the inflaton $\phi$. Their back reaction
can be traced through their effect in the equation of motion of the
inflaton. Strictly speaking the back reaction effects we want to
consider arise from the one loop correction to the inflaton
potential. Noting the action (\ref{S2}) and the form of the
potential $V_0$, these corrections are%
\be\label{back-reacted-phi}%
\ddot\phi+3H\dot\phi+V'_0+ \Delta V'=0\ , %
\ee%
where \footnote{Note that at the one loop level besides  $\Delta V'$
terms one should consider the running of the coupling constants
$\lambda_{eff}$, $\kappa_{eff}$ and $m^2$. Since we are dealing with
quadratic potentials which are renormalizable, one may compute these
loop corrections. Due to the large number of fields these one loop
effects could be large. In our case, however, we can just use the
renormalized values for these parameters. We also note that among
the specific cases that we discussed, the $\lambda\phi^4$ theory at
one loop level will receive a contribution of the form $\delta
m^2\phi^2$. The condition of having an inflection point,
$\kappa^2_{eff}=m^2\lambda_{eff}$ will not be preserved by the
quantum corrections (at one loop level). The symmetry breaking
potential, however, will preserve its form. Besides the
renormalization of the couplings there is also the Coleman-Weinberg
corrections. In our discussion, however, we will not consider this.
The symmetry breaking case can be viewed as the bosonic part of the
potential in a supersymmetric theory, for which the Coleman-Weinberg
correction is absent.%
}%
\be%
\Delta V'=3\lambda_{eff} \phi \langle \phi^2\rangle
-2\kappa_{eff}\langle \phi^2\rangle+ \frac12\sum_{r,lm} \langle
\Psi^\star_{r,lm}\Psi_{r,lm}\rangle \frac{d{M_{r, l}^2(\phi)}}{{d\phi}}%
\ee%
where $\langle \Psi^\star_{r,lm}\Psi_{r,lm}\rangle$ during inflation
should be replaced with  $N_\Psi$ \eqref{Npsi-result}. The $\langle
\phi^2\rangle$ during slow-roll inflation where $\epsilon, \eta$ are
very slowly varying (and are basically constant), is negligible and
can be dropped.\footnote{The $\langle \phi^2\rangle$ is essentially
the same $\langle\Psi\Psi\rangle$ as the zero $\Psi$ modes for which
the $\nu$ factor, which causes the enhancement in the $\alpha$ and
$\beta$ modes of $\Psi$, is absent. Therefore, the $\langle
\phi^2\rangle$ is negligible compared to $\langle
\Psi_{r,lm}^2\rangle$ term.} This term, however, may become large
toward the end of inflation when both $\epsilon$ and
$\dot\epsilon/H$ can become order one. This is the regime we
consider in the next subsection. During the slow-roll inflation
therefore, $\Delta V'=\frac12 N_\Psi dM^2(\phi)/d\phi$.

Let us focus on the $\lambda_{eff}\phi^4/4$ theory for which the
computations of \ref{infpsi} was mainly carried
out. In this case the back-reaction equation of motion for the inflaton $\phi$, is%
\be\label{back-react-phi4-theory}%
\ddot\phi+3H\dot\phi+\lambda_{eff}\phi^3+ \nu \lambda_{eff} N_\Psi\ \phi=0\ . %
\ee%
As the computation of $N_\Psi$ was carried out assuming the slow-roll
approximation is valid, we stress that the above equation is hence
only trustable during slow-roll inflation.

To check whether the last term is harmful to the (slow-roll)
inflationary dynamics we compare the last term to contribution of
the main potential  driving inflation:
\be%
%\frac{\frac12 \frac{dM^2}{d\phi} N_\psi}{V_0'}=%
\frac{\lambda_{eff} \, \nu \phi\cdot N_\psi} {\lambda_{eff} \phi^3}=
\frac{27}{128\pi^2}\ \left(\frac{H}{M_{P}}\right)^2\cdot
\epsilon^{5/4}\
\delta^{1/2}e^{-\pi\delta}\ .%
\ee%
Recalling the WMAP bound $H/M_{P}< 10^{-4}$, the above expression is
of course much smaller than one during slow-roll inflation. We
therefore conclude that the particle creation is not going to
destroy our M-flation model.

%%%%%%%%%%%%%%%%%%%%%%%%%%%%%%%%%%%%%%%%%%%%%%
\subsection{Particle creation and the preheat
scenario}\label{preheat-subsection}

We argued that during the slow-roll inflation particle creation, and
hence its back-reaction on the dynamics of the $\phi$ field is not
large. However, particle creation can
become important when $\epsilon, \eta$ are of order one. In this
section we explore this region. The equations we employ are of
course the equation of motion for $\Psi_{r,lm}$ modes (\ref{KG}) and
the modified equation for the inflaton field
\eqref{back-reacted-phi}.

We follow the line of analysis performed in \cite{Greene-Kofman}.
That is, in first step we ignore the $\Delta V$ term in
\eqref{back-reacted-phi}, i.e. we study classical dynamics of the
inflaton field $\phi$ when $\epsilon$ is of order one and then study
the dynamics of the $\Psi_{r,lm}$ fields using this solution for $\phi$ as
background, and finally, we  include $\Delta V$ in the equation of $\phi$.
The inflaton potential we start with \eqref{Vphi} is a generic
quartic potential and the analytic treatment of the equations is not
possible for generic values of parameters $\lambda_{eff},\
\kappa_{eff},\ m^2$. The analysis for the chaotic case, i.e. when
only $\lambda_{eff}$ (or when $m^2$) is non-zero has been carried
out in some detail in \cite{Greene-Kofman, preheating-II,
Kaiser:1997mp, Tsujikawa:2002nf}. In our model, however, the case with only non-zero
$m^2$ does not have the quartic coupling to the preheat field(s) and
hence does not involve  a preheating model. We therefore focus on
the $\lambda\phi^4/4$ theory.

For the $\lambda\phi^4/4$ theory, the  potential for the fields is%
\ba\label{potential-phi4}%
%\begin{split}%
V(\phi,\Psi_{r,lm})&=&\frac14\lambda_{eff}\phi^4+
\frac12\lambda_{eff}\phi^2 \sum_{r,lm} \frac12(\omega^2-\omega) \Psi^\star_{r,lm}\Psi_{r,lm} \\ %
&=&\frac14\lambda_{eff}\phi^4+ \frac12\lambda_{eff}\phi^2
\sum_{m=1}^{N^2-1} |\Psi_{0m}|{}^2 \nonumber\\
&+& \frac12\lambda_{eff}\phi^2 \left[ \sum_{l=0}^{N-2}
\frac{(l+3)(l+2)}{2} \sum_{m=1}^{2l+1} |\Psi_{\alpha,\ lm}|{}^2
+\sum_{l=1}^{N} \frac{(l-2)(l-1)}{2} \sum_{m=1}^{2l+1}
|\Psi_{\beta,\ lm}|{}^2 \right]  \nonumber
%\end{split}%
\ea%
where  in the second line we have decomposed $\Psi_{r,lm}$ into
zero, $\alpha$ and $\beta$ modes for which $\omega$ respectively
takes values $-1$, $-(l+2)$ and $l-1$. We remark that the potential
\eqref{potential-phi4} is an approximation to the potential term we
start with \eqref{action} (for the  $\kappa=0,\ m^2=0$ case) to
order $\Psi^2$.

As shown in \cite{Greene-Kofman}, the $\phi$ equation of motion
toward the end of inflation is an (anharmonic) oscillator around
$\phi=0$, the amplitude of the solution is decreasing as
$t^{-1/2}$. In this regime
 the $\phi$ field is effectively living in a radiation dominated
background with $1/H=2t$. It appears that the equation for $\phi$
takes a simple form in the conformal time, with Jacobi (elliptic)
cosine function as its solution \cite{Greene-Kofman}. The equation
for the $\Psi$ modes, too, can be solved explicitly in our case. In
fact in the notations of \cite{Greene-Kofman}, the equation for all
three zero, $\alpha$ and $\beta$ modes is of the form of
$g^2/\lambda=n(n+1)/2$ (with $n=1, l+2, l-2$ respectively for zero,
$\alpha$ and $\beta$ modes) for which most of the calculations can
be performed analytically. (Note that $g^2/\lambda$ of
\cite{Greene-Kofman} coincides with the parameter $\nu$
\eqref{M2-phi4} in our model.) As discussed in \cite{Greene-Kofman}
for these specific values of $g^2/\lambda$ we have the significant
property that there is an enhancement in the parametric resonance
leading to considerable creation of zero, $\alpha$ and $\beta$
modes. As discussed in \cite{Greene-Kofman} one can distinguish two
even and odd $n$ cases. For the odd $n$ (i.e. for our zero modes,
odd $l$ $\alpha$ and $\beta$-modes) the particle creation is peaked
around zero momentum $k$ modes. For the even $n$ modes, however, the
particle creations is peaked around momenta $k^2=\frac32
H_{inf}^2\epsilon \sqrt{\frac{g^2}{2\lambda}}$, where $\epsilon$ is
the computed for the beginning of the slow-roll inflation and
$H_{inf}$ is the Hubble during inflation. For low $n$, the  Floquet
index $\mu_k\propto \ln n_k$ ($n_k$ is the number density of the
produced particles at momentum $k$) is around $0.15$ for odd $n$ and
around $0.05$ for even $n$. Therefore, among the low $n$ modes the
main contribution to preheating is coming from odd $n$
\cite{Greene-Kofman}. As discussed in \cite{Greene-Kofman} the
bigger $k$ is, the more energy can be transferred from the
inflationary sector to the $\Psi$ sector and a more efficient
preheat mechanism. This means that $\alpha$ and $\beta$ modes with
large $l$, $l$ of order $N$, make the biggest contribution, this is
despite the fact that the zero modes have a larger degeneracy (of
order $N^2$) compared to the degeneracy of order $N$ for the large
$l$ modes. All in all, due to the existence of the large $l$ modes,
and for large $N$ in our model we expect to have a very efficient
preheating model. The computations for the modes with large
$g^2/\lambda$ has been carried out in \cite{Greene-Kofman} and the
only point which is different in our case is that their result
should be multiplied with the degeneracy factor $2l+1$.

%We have not provided a mechanism of reheating to transfer energy
%from the $\Psi_{r,lm}$ fields to the Standard Model of particle
%physics.

As we argued we have an efficient preheat mechanism in our model. As
a very crude estimate of the preheat temperature in our M-flation
setup we may hence use an instant efficient preheating that all the
energy of the inflaton field has gone to effectively massless zero
modes by the end of inflation, leading to%
\ba
  N^2 T^4 \sim 3 H^2 M^2_{P} \, ,
  \ea
where  $N^2$ estimates the number of species and $T$ is the preheat
temperature. As we see, this is as if we have effectively an instant
preheating model in which the maximum temperature achieved is
lowered by $1/\sqrt{N}$. As a rough estimate taking $H$ saturating
its current bound $H\sim 10^{-5} M_{P}$ and $N\sim 10^5$ then the
preheat temperature becomes of order $T\sim 10^{13}$ Gev. Reducing
the preheat temperature to below GUT scale is in principle a
positive feature, as it removes the problem with overproduction of
gravitinos.

%%%%%%%%%%%%%%%%%%%%%%%%%%%%%%%%%%%%%%%%%%%%%%%%%%%%%%%%%%%%%%%%%%%%%%%%%%%%%%%%%%%%%%%%%%%%

\section{Motivation from String Theory}
\label{stringtheory}

Here we argue that our M-flation setup presented in section
\ref{setup-section} with non-commutative matrices and potential in
the form of (\ref{The-Potential}) is strongly motivated from string
theory.

In the context of string theory, the world-volume theory of $N$
coincident p-branes is described by a (supersymmetric) $U(N)$ gauge
theory. In this system, the transverse positions of the branes,
$\Phi_{I},\ I=p+1,...,9$, which from the world-volume theory are
scalars in the adjoint representation of $U(N)$,  are hence $N\times
N$ matrices. For the case of our interest, $p=3$, there are six such
scalars. The DBI action for the system of $N$ coincident D3-branes
 in the background RR six form flux (sourced by a distribution of
 D5-branes) is given by (e.g. see \cite{Myers:1999ps})%
 \ba \label{DBI}%
 S= \frac{1}{(2\pi)^3 l_s^4g_s} \int d^{4}x \, \STr \left( 1-\sqrt {-|g_{ab}| } \sqrt { |Q^{I}_{J} |}\,
+\frac{i g_s}{2\cdot 2\pi l_s^2} [X^{I},
X^{J} ] C^{(6)}_{I\, J\, 0 1 2 3 } \right)%
\ea%
Here $l_s$ is the string scale and $g_{s}$ is the perturbative
string coupling. The operator $\STr$ on a product of matrices is the
trace of their symmetrized product. The induced metric on branes,
$g_{ab}$, is given by $g_{ab} = G_{MN} \partial_a X^{M}
\partial_b X^{N}$, where $X^{M}$ indicates the ten-dimensional
positions of the branes and $G_{MN}$ is the ten dimensional
background metric. Here the indices $I, J=4,5,\cdots, 9$ represent
the coordinates perpendicular to the branes world-volume, the
indices $a, b=0,1,2,3$ represent the brane world-volume coordinates
and the capital letters $M, N=0,1,\cdots, 9$ indicate the
ten-dimensional coordinates. The matrix $Q^{I}_{J}$ is due to
non-commutativity
properties of the system  given by%
\ba%
Q^{IJ} = \delta^{IJ} + \frac{i}{2\pi l_s^2} [ X^{I}, X^{J}] \, ,%
\ea%
%where $\lambda= 2\pi l_{s}^{2}$ and we have used the convention such
%that $X^{I}= \lambda \hat \Phi^{I}$, so the scalars $\hat\Phi^{I}$
%have dimension of mass.%
and  $C^{(6)}_{I\, J\, 0 1 2 3 }$ is a rank-6 antisymmetric
Ramond-Ramond (RR) field which has  two legs along the direction
transverse to the D3-brane.

We consider the ten-dimensional IIB supergravity background%
\be\label{sugra-background}%
\begin{split}%
ds^2 &=-2dx^+dx^--\hat m^2 \sum_{i=1}^3 (x^i)^2 (dx^
+)^2+\sum_{I=1}^8d
x_I dx_I \\
C_{+123ij}&= \frac{2\hat\kappa}{3} \epsilon_{ijk} x^k
\end{split}%
\ee%
where $i,j$ indices, which are ranging over $1,2,3$, parameterize
three out of six transverse directions to D3-brane and $x^I$ include
three spatial directions along the brane and five of the transverse
directions to D3-branes. With $\hat m^2=4g_s^2 \hat\kappa^2/9$ the
above background, with constant dilaton, is a solution to
supergravity equations of motion. This background is very similar to
the background of  \cite{Polch-Srass} (see also Appendix D of
\cite{M5-brane} for a discussion on the Matrix model on the above
background).

If we turn-off fluctuations along the directions transverse to the
branes and the $x^i$ directions (this may be done if we compactify
these three directions on a $T^3$ of very small radius), fix the
light-cone gauge on the D3-branes, expand the action and keep up
to order four in $X^I$, we obtain%
\ba%
S&=& \frac{1}{(2\pi)^3 l_s^4g_s} \int d^4x\ \Tr\bigl[\frac{-1}{2}
\partial_{\mu} X_{i}\partial^{\mu} X_{i} - V(X)\bigr]\cr \;&\;&\; \\
V&=& - \frac{1}{4\cdot(2\pi l_s^2)^2}\ [ X_{i}, X_{j}] [ X_{i},
X_{j}]
 +\frac{i g_s\hat \kappa}{3\cdot 2\pi l_s^2} \epsilon^{ijk} X_i [X_{j}, X_{k} ]
 X_{i}+\frac{1}{2} \hat m^2 X_i^2\ .\nonumber
\ea%
If we redefine%
\be%
X_i=\sqrt{(2\pi)^3 g_s}\ l_s^2 \Phi_i%
\ee%
and upon addition of the four-dimensional Einstein gravity, the
above action takes the form of \eqref{action} with the potential
\eqref{The-Potential} once we identify the parameters as
\be%
\lambda=2\pi g_s\ ,\qquad \hat\kappa= \frac{\kappa}{g_s\cdot
\sqrt{2\pi
g_s}}\ ,\qquad \hat m^2=m^2\ . %
\ee%

Although from the brane theory viewpoint we need to choose $\hat
m^2$ and $\hat \kappa$ such that \eqref{sugra-background} is a
solution to supergravity, namely $\lambda m^2=4\kappa^2/9$, since we
presented the brane theory by the way of motivation, at the level of
M-flation action we may relax this condition and take $\lambda,\
\kappa$ and $m^2$ as independent parameters. It is also worth noting
that $\lambda m^2=4\kappa^2/9$ corresponds to the ``symmetry
breaking'' inflation potential (\ref{Vsusy}). Furthermore, the
minimum $\phi=\mu$ for the potential (\ref{Vsusy}) corresponds to
the supersymmetric background where $N$ D3-branes blow-up into a
giant D5-brane.

It is worth noting that in the brane theory setting the $U(N)$
symmetry appears as a gauge symmetry, while in our model we took it
to be a global symmetry. Promoting $U(N)$ to a gauge symmetry does
not change our analysis of the $SU(2)$ sector and the corresponding
inflationary dynamics. Due to the gauge symmetry, however, not all
the $\Psi_i$ modes are physical. Among them the zero modes can be
removed by the gauge transformations and hence in the theory where
$U(N)$ is gauged only we deal with $\alpha$ and $\beta$ modes. Thus
in the gauge symmetry case, from the first $N^2+1$ iso-curvature
modes in the mass tower, only one ($l=0$ $\alpha-$mode) remains,
which has an amplitude of order few percent of curvature spectrum.
However this will not change the analysis of section
\ref{not-harmful-section}, as the main contribution were coming from
$\alpha$ or $\beta$ modes of $l$ in the window \eqref{nu-window}
which are also present in the gauged theory. The analysis of
preheating mechanism of section \ref{preheating-section} will remain
valid because again zero modes do not have the main contribution.

As a result of motion of D3-branes in the background $C^{(6)}$ flux,
two of the directions transverse to D3-branes blow-up into (fuzzy)
two sphere, which in the large $N$ limit behaves as a D5-brane with
world-volume $R^4\times S^2$ \cite{Myers:1999ps}. In this geometric
picture our inflaton field $\phi$ is nothing but the radius of this
two-sphere. In this sense the effective inflaton field $\phi$ in our
M-flation scenario is closely related to the inflaton in the ``giant
inflaton'' model of \cite{giant-inflaton} (see also
\cite{Ward:2007gs}).

%%%%%%%%%%%%%%%%%%%%%%%%%%%%%%%%%%%%%%%%%%%%%%
\section{Discussion}

In this work we have presented a new inflationary scenario, the
M-flation, in which inflation is driven by matrix valued scalar
fields. M-flation, hence, falls into the general class of
multi-field inflation models and shares positive features of
N-flation. Specifically, we used M-flation to obtain a super-Planckian (large field) field variation
during inflation. This leads to a
considerable amount of gravity waves which can be
detected in future gravity wave observations such as  PLANCK \cite{PLANCK, Efstathiou:2009xv}, CMPOL \cite{Baumann:2008aq} and QUIET \cite{quiet}.
Moreover, as we discussed, due to the scalings with  powers of $N$,
the dimension of the matrices, M-flation bears a solution to the
fine-tuning problem of the coupling in the $\lambda\phi^4/4$ chaotic
inflation.

Here we focused on a special class of M-flation scenarios with
potentials of the form \eqref{The-Potential} where the potential is
quadratic in powers of $\Phi_{i}$ and their commutators. Within our
three parameter family of the potentials there are  interesting
special models of inflation which were analyzed in section
\ref{Inflation-in-M-flation}. One may, however, start with other
forms for the potential. Specifically, if one starts from the string
theory realization of the scenario, then from action (\ref{DBI}) one
obtains higher power corrections in terms of $\Phi_{i}$ and
$[\Phi_{i}, \Phi_{j}]$. Furthermore, the kinetic energy may also
have a non-trivial form and one may combine the idea of M-flation
with a DBI non-standard kinetic energy \cite{Alishahiha:2004eh}. It would be interesting to
see the spectrum of the adiabatic and iso-curvature perturbations
for this case of ``DBI M-flation''.

 One of the observable effects of multi-field inflation models
is that besides the usual power spectrum of the adiabatic
fluctuations $P_{\cal R}$, we also have a non-zero power spectrum
for the iso-curvature perturbation, $P_{ {\cal S}_{r,l m} }$. We
analyzed the ratio $P_{ {\cal S}_{r,l m} }  / P_{\cal R} $ for
various inflationary models up to the end of inflation. Inside the
Hubble radius this ratio is close to unity but once the mode leaves
the Hubble radius this ratio decays quickly towards the end of
inflation. As shown, our analytical estimates of   $P_{ {\cal
S}_{r,l m} } / P_{\cal R} $  are in good agreement with the
numerical results. In order to relate this ratio to the observed
values from CMB, however, we should also supplement our M-flation
scenario with a reheating mechanism.

As we discussed in section \ref{iso}, the iso-curvature
perturbations $\Psi_{r,lm}$ do not produce entropy perturbations nor
couple to curvature perturbations. This is due to our initial
conditions resulting in the fact that they are classically frozen
during inflation. This in turn implies that they do not carry energy
up to leading order in perturbation theory. However, they can
contribute to entropy perturbations at second order in perturbation
theory or through preheating mechanism. Via the same mechanisms, the
iso-curvature perturbations $\Psi_{r,lm}$ can produce
non-Gaussinities which are under intense observational
investigations.

%The observed value of $P_S/P_R$ on CMB not
%only depends on its primordial value during inflation, but also to the reheating mechanism.
As was discussed in section
\ref{preheating-section} our model naturally contains a preheating
sector (essentially the $\Psi_{r,lm}$-modes) which due to the large
number of these modes works very efficiently, taking away the energy
stored in the inflaton field. In order to complete our model, we
need to have a reheating model via which the energy of the
$\Psi$-modes is transferred into the Standard Model particles. This
is postponed to future works.
As a possibility for reheating mechanism in string theory setup,
where $\Phi_{i}$ represents the collective positions of $N$
D3-branes, we may imagine that the Standard Model of particle
physics are confined to the branes in the forms of open strings
gauge fields $A^{(a)}_{\mu}$. The question of reheating would be how
to transfer energy from the $\Phi_{i}$ fields, more precisely from
the $\Psi_{r,lm}$ modes, to the open string gauge fields
$A^{(a)}_{\mu}$.

In this work we have restricted the analysis to a particular
solution where $\Psi_{r,lm}$ fields are absent in classical
inflationary dynamics. The inflaton field  $\phi$ is the projection
of $\Phi_{i}$ along the $N\times N$ irreducible representation of
$SU(2)$, the $J_{i}$ matrices. In the field space of
$\phi-\Psi_{r,lm} $ this corresponds to a straight inflationary
background. In general, one may consider an arbitrary initial
condition where $\Psi_{r,lm}$ fields are turned on. In the field
space of $\phi-\Psi_{r,lm} $ this gives a complicated curved
inflationary trajectory. One such possibility with a more controlled
dynamics is to take $J_i$ to form a reducible $N\times N$
representation of $SU(2)$ which consists of $n$ irreducible blocks.
In this case the classical inflationary dynamics of our theory
reduces to that of $n$ decoupled scalar fields, each with generic
quartic potential. In this case the iso-curvature perturbations
would be non-adiabatic and a significant amount of entropy
perturbations can be created during inflation. Similarly, one
expects a considerable amount of non-Gaussianities to be produced in
this case. It would be interesting to build an specific model of
M-flation where $\Psi_{r,lm}$ fields are turned on during inflation
and calculate non-Gaussianities and entropy perturbations produced
and compare them with the observational bounds.

\vspace{0.5cm}

%%%%%%%%%%%%%%%%%%%%%%%%%%%%%%%%%%%%%%%%%%%%%%
\noindent {\bf \large {Acknowledgment}}

%\vspace{5mm}

We thank Y. Farzan, N. Khosravi and K. Nozari   for useful
discussions. We thank K. Turzynski for some computational
assistance. We specially thank B. Bassett and  R. Brandenberger for
comments on the draft and for many useful insights. H. F. would like
to thank KITPC, and M.M.Sh-J  the Abdus-Salam ICTP, for hospitality
during the final stage of this work. A. A. is supported by NSERC of
Canada and MCTP.

\appendix

%%%%%%%%%%%%%%%%%%%%%%%%%%%%%%%%%%%%%%%%%%%%%%%%%%%%%%%%%%%%%%%%%%%%%%%%%%%%%%%%%%%%%%%%%%%%
\section{ Symmetry breaking  inflation}\label{Appen-A}

Here we study inflation from the symmetry breaking potential in some
details. Suppose inflation starts when $ \phi_{i}> \mu$. The total
number of e-folds $N_{e}$ is obtained by \ba \label{Ne}
M_{P}^{2}\, N_{e}& = & \int_{\phi_{f}}^{\phi_{}{i}} \frac{d \phi \, V}{V'}  \nonumber\\
&=& \frac{1}{8} (y-x) - \frac{\mu^{2}}{32} \ln \left( \frac{\mu^{2} + 4 y}{ \mu^{2} + 4 x  }
 \right)
%&=& \frac{1}{8} \psi_{i}(\psi_{i} - \mu) - \frac{1}{8} \psi_{f}(\psi_{f} - \mu)
%- \frac{\mu^{2}}{32} \ln \left( \frac{    \mu^{2} + 4 \psi_{i} (\psi_{i} - \mu)   }{   \mu^{2} + 4 \psi_{f} (\psi_{f} - \mu)   }   \right) \, .
\ea where $\phi_{f}$ is the endpoint of inflation and for later
convenience we have defined $x=\phi_{f}(\phi_{f} - \mu)$ and $ y=
\phi_{i}(\phi_{i} - \mu) $. As usual, define the slow-roll
parameters \ba \epsilon= \frac{1}{2} M_{P}^{2}\left( \frac{V'}{V}
\right)^{2} \quad , \quad \eta=  M_{P}^{2} \frac{V''}{V}  \, . \ea
where $'$ denotes derivative with respect to $\phi$. Inflations ends
when $\epsilon=1$, which is used to fix  $x$ and $\phi_{f}$ \ba x  =
4 M_{P}^{2} + M_{P} \sqrt{  16 M_{P}^{2} + 2 M_{P}^{2} \mu^{2}     }
\ea The scalar spectral index at $\phi_{i}$  is $n_{{\cal
R}}-1=2\eta - 6 \epsilon$  at $\phi_{i}$   which can be used to
eliminate $y$ \ba \frac{y}{M_{P}^{2}} =\frac{12+ \sqrt{  144 + 8
(1-n_{{\cal R}}) \frac{\mu^{2}}{M_{P}^{2}}    } }{ 1-n_{{\cal R}}  }
\ea

Plugging these values for $x$ and $y$ in (\ref{Ne}), we find an
equation for $\mu/M_{P}$. Solving this equation numerically  with
$N_{e}=60$ and $n_{s}=0.96$ from WMAP5 central value, one obtains
$\mu/M_{P} \sim 26$. This in turn yields $\phi_{i} \simeq 44 M_{P}$
and $\phi_{f}\simeq 28 M_{P}$.

The COBE normalization, can be used to fix the value of $\lambda_{eff}$
\ba
\delta_{H} = \frac{1}{\sqrt {75} \pi} \frac{V^{3/2}}{M_{P}^{3} V'}
= \frac{\lambda_{eff}^{1/2}}{4 \sqrt{75} \pi } \frac{ \phi_{i}^{2} (\phi_{i} - \mu)^{2}}{  ( 2 \phi_{i} - \mu) M_{P}^{3}  } \, .
\ea
Using $\delta_{H} \simeq 2 \times 10^{-5}$ and the above values for $\phi_{i}$
and $\mu$, one obtains $\lambda_{eff} \simeq 10^{-14}$. This corresponds to
$N \sim 10^{5}$ as in chaotic inflation case.

It it is also instructive to look into gravity wave amplitudes,
determined by  quantity $r$, defined as the ratio of gravitational
perturbation amplitude to scalar perturbation amplitude at
$\phi_{i}$: \ba r&=& \frac{8}{3} (1-n_{{\cal
R}}) + \frac{16}{3} M_{P}^{2} \frac{V''}{V}\nonumber\\
&=& 4 (1-n_{{\cal R}} ) + 32 \frac{M_{P}^{2}}{y} \, . \ea Using the
above values for $\phi_{i}$ and $\mu$, one obtains $r\lesssim 0.2$
which is consistent with the upper bound $r<0.22$  from WMAP5.

The analysis when inflation takes place in regions $ \mu/2 <
\phi_{i}<\mu$ and $0< \phi_{i} < \mu/2$ is similar to the above.

%%%%%%%%%%%%%%%%%%%%%%%%%%%%%%%%%%%%%%%%%%%%%%%%%%%%%%%%%%%%%%%%%%%%%%%%%%%%%%%%%%%%%%%%%%%%

%\section*{References}

\end{document}